\documentclass[twocolumn,pra,superscriptaddress,showpacs,aps,groupedaddress,floatfix]{revtex4-1}
\usepackage{graphicx}
\usepackage{dcolumn}
\usepackage{stmaryrd}
\usepackage{latexsym}
\usepackage{amssymb}
\usepackage{amsfonts}
\usepackage{amsmath}
\usepackage{mathtools}
\usepackage{fancybox}
\usepackage{color}
\usepackage{bigdelim}
\usepackage[breaklinks=true,colorlinks,citecolor=blue,linkcolor=blue,urlcolor=blue]{hyperref}
\usepackage{float}
\usepackage[capitalize]{cleveref}

\usepackage{epstopdf}
\usepackage{braket}

\usepackage{graphicx}
\usepackage[dvipsnames]{xcolor}
\usepackage[normalem]{ulem}

\usepackage[straightquotes]{newtxtt}

\usepackage{lmodern}
\newcommand{\pv}{\mathbf{p}}
\newcommand{\qv}{\mathbf{q}}
\newcommand{\rv}{\mathbf{r}}
\newcommand{\zerov}{\mathbf{0}}
\newcommand{\veck}{\mathbf{k}}

\newcommand{\nablav}{\boldsymbol \nabla}
\newcommand{\Ac}{\mathcal{A}}
\newcommand{\Zc}{\mathcal{Z}}
\newcommand{\nnl}{\nonumber \\}
\newcommand{\Aup}{A_{\psi,k}}
\newcommand{\mup}{m_{\psi,k}}
\newcommand{\Ad}{A_{\phi,k}}
\newcommand{\md}{m_{\phi,k}}
\newcommand{\Aphi}{A_{t,k}}
\newcommand{\mphi}{m_{t,k}}
\usepackage{cleveref}
\DeclareMathOperator{\STr}{STr}

\begin{document}
\title{Functional-renormalization-group approach to strongly coupled Bose-Fermi mixtures in two dimensions}
\author{Jonas von Milczewski}
\affiliation{Max-Planck-Institute of Quantum Optics,
	Hans-Kopfermann-Stra{\ss}e 1, 85748 Garching, Germany}
\affiliation{Munich Center for Quantum Science and Technology (MCQST),
	Schellingstra{\ss}e 4, 80799 Munich, Germany
}
\author{F\'{e}lix Rose}
\affiliation{Max-Planck-Institute of Quantum Optics,
	Hans-Kopfermann-Stra{\ss}e 1, 85748 Garching, Germany}
\affiliation{Munich Center for Quantum Science and Technology (MCQST),
	Schellingstra{\ss}e 4, 80799 Munich, Germany
}
\author{Richard Schmidt}
\affiliation{Max-Planck-Institute of Quantum Optics,
	Hans-Kopfermann-Stra{\ss}e 1, 85748 Garching, Germany}
\affiliation{Munich Center for Quantum Science and Technology (MCQST),
	Schellingstra{\ss}e 4, 80799 Munich, Germany
}
\date{\today}
\begin{abstract}

We study theoretically the phase diagram of strongly coupled two-dimensional Bose-Fermi mixtures interacting with attractive short-range potentials as a function of the particle densities. We focus on the limit where the size of the bound state between a boson and a fermion is small compared to the average interboson separation and develop a functional-renormalization-group approach that accounts for the bound-state physics arising from the extended Fr\"{o}hlich Hamiltonian. By including three-body correlations we are able to reproduce the polaron-to-molecule transition in two-dimensional Fermi gases in the extreme limit of vanishing boson density. We predict frequency- and momentum-resolved spectral functions and study the impact of three-body correlations on quasiparticle properties. At finite boson density, we find that when the bound-state energy exceeds the Fermi energy by a critical value, the fermions and bosons can form a fermionic composite with a well-defined Fermi surface. These composites constitute a Fermi sea of dressed Feshbach molecules in the case of ultracold atoms while in the case of atomically thin semiconductors a trion liquid emerges. As the boson density is increased further, the effective energy gap of the composites decreases, leading to a transition into a strongly correlated phase where polarons are hybridized with molecular degrees of freedom. 
We highlight the universal connection between two-dimensional semiconductors and ultracold atoms and we discuss perspectives for further exploring the rich structure of strongly coupled Bose-Fermi mixtures in these complementary systems.
\end{abstract}

\maketitle

\section{Introduction}
\label{sect:Intro}
Ever since the  theoretical explanation of conventional superconductivity as arising from the attractive interaction between electrons mediated by phonons~\cite{Bardeen1957,Bardeen1957a}, Bose-Fermi mixtures have been the subject of intense research. As they combine systems of different quantum statistics, their many-body behavior can be vastly different from that of the underlying bosonic or fermionic subsystems alone. Consequently, they can feature rich many-body physics ranging from superconductivity to the formation of composite bosonic or fermionic bound states similar to mesons and baryons in particle physics.

In solid-state physics, bosons typically appear as collective degrees of freedom. These may be, for instance, phonon excitations of an underlying crystalline lattice or  collective excitations of the electronic system itself in the form of, e.g., magnons or plasmons. Beyond such systems, experimental progress in the fields of atomically thin semiconductors \cite{Mak2013} and ultracold atoms \cite{Chin2010} makes it now possible  to enter a new regime of \textit{strongly coupled} Bose-Fermi mixtures. Here ---akin to the physics of nuclear matter--- fermions and bosons appear on equal footing, both representing pointlike particle degrees of freedom. 

Crucially, direct pairing between bosons and fermions is a  new essential ingredient in these mixtures. Recently it was shown \cite{rath2013} that for such strongly coupled Bose-Fermi mixtures a description in terms of Fr\"ohlich or Holstein models \cite{Froehlich1954,Holstein1959}, in which fermions couple linearly to the bosonic degrees of freedom, fails. In addition,  the coupling to bosons at quadratic order  becomes relevant, which has to be accounted for in an \textit{extended} Fr\"ohlich Hamiltonian \cite{rath2013}, giving rise to qualitatively new physics recently observed in experiments in cold gases \cite{Hu2016,jorgensen2016,yan2020bose} and Rydberg systems \cite{Camargo2018}. 

 Various aspects of  atomic, three-dimensional Bose-Fermi mixtures  have been investigated theoretically using  the Fr\"ohlich model ---thus disregarding the crucial quartic interaction term. This revealed a rich structure of the phase diagram ranging from polaron formation \cite{Tempere2009,Casteels2011,Shashi2014} and boson-induced $p$-wave superfluidity \cite{Kinnunen2018}, to phonon softening and phase separation \cite{Enss2009}.

Similarly, the phase diagram of \textit{two-dimensional} Bose-Fermi mixtures has been explored using the Fr\"ohlich model. These studies were  motivated in particular by exciton-electron mixtures in semiconductors, and, following initial work by Ginzburg \cite{Ginzburg1964}, it was predicted that the system may turn superconducting \cite{Laussy2010,Laussy2012,Cherotchenko2016} while other works proposed a transition to supersolidity \cite{Shelykh2010,Matuszewski2012}, or that the formation of both phases might be intertwined \cite{Cotlet2016}. 

Due to the shortcomings of the Fr\"ohlich model and mean-field inspired approaches that neglect pairing \cite{Viverit2000,Efremov2002,Roth2002,Roth2002a,Albus2002,Hu2003}, these initial studies  missed the fact that the microscopic interaction between atoms in ultracold gases and between excitons and electrons in semiconductors is fundamentally attractive. While in cold gases  interactions arise from long-range van der Waals forces,  the polarization of charge-neutral excitons by electrons gives rise to attractive forces in semiconductors. Crucially, in both cases the interactions support bound states between the fermionic and bosonic particles. Consequently, as the strongly coupled regime is entered, one has to consider the  \textit{extended} Fr\"ohlich Hamiltonian in order to account for the  pairing  to  fermionic Feshbach molecules in cold atoms and  exciton-electron bound states, called trions, in semiconductors. 

\begin{figure}[t]
	\begin{center}
		\includegraphics[width=\linewidth]{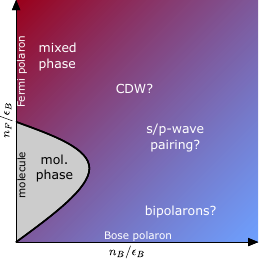}
	\end{center}
	\caption{Schematic phase diagram of  two-dimensional Bose-Fermi mixtures as a function of the  density of either species.  At strong-coupling the system is described by the extended Fr\"ohlich model that accounts for the formation of a two-body bound state between fermions and bosons of energy $\epsilon_B$. The limit $n_B=0$ (along the y-axis) defines the Fermi polaron problem, discussed in \cref{sect:polaron}, where a single bosonic impurity interacts with  a fermionic bath. In this limit the impurity can either bind with a fermion into a  molecule  or remain unbound as a Fermi polaron.
	At finite boson density, discussed in \cref{sect:finitedens},  we find a transition from a molecular  phase   which hosts a Fermi sea of bound molecules (light gray) to a mixed phase in which a condensate of bosons hybridizes fermionic and molecular degrees of freedom (red/dark shading).  With the exception of the extreme limit $n_F=0$ that corresponds to the Bose polaron problem (along the x-axis), as the boson density is increased beyond the regime $n_B\ll n_F$  (red/dark to blue/light shading),  the phase diagram remains largely unexplored. Starting with the possibility of bipolaron formation \cite{camacho2018}, various competing phases can be conjectured based on studies of the simpler weak-coupling Fr\"ohlich model, ranging from supersolid charge density wave states \cite{Shelykh2010,Cotlet2016} to boson-mediated $s/p$-wave fermion pairing \cite{Enss2009,Laussy2010,Laussy2012,Kinnunen2018}. } \label{SchematicPDintro} \end{figure}

The presence of this novel bound-state physics  renders the description of strongly coupled Bose-Fermi mixtures an outstanding theoretical challenge. This is reflected by the fact that until now ---except for initial studies in three dimensions \cite{Powell2005,Watanabe2008,Fratini2010,Ludwig2011,Yu2011,Bertaina2013,Guidini2014,Guidini2015}--- the phase diagram  of strongly coupled Bose-Fermi mixtures as function of the density of bosons $n_B$ and fermions  $n_F$, schematically shown in Fig.~\ref{SchematicPDintro}, remains unexplored.  With the discovery of atomically thin transition-metal-dichalcogenides, the semiconducting class of layered van der Waals materials, the exploration of this phase diagram in two dimensions becomes particularly urgent.  This is  not only due to the potential of layered materials for technological applications, but also due to the possibility of realizing long-lived, stable exciton-electron mixtures that feature a striking similarity to  cold atomic mixtures \cite{Fey2020,Imamoglu2021,Efimkin2020}. This  universal connection, detailed by the comparison  of typical scales in both systems shown in Table~\ref{tab:1} below, opens the possibility to explore emerging phases in strongly interacting systems in two complementary and seemingly  disparate systems, that while playing on vastly different energy and length scales, are governed by the same dimensionless system parameters.

In this work, we study theoretically the phase diagram of strongly coupled two-dimensional Bose-Fermi mixtures as a function of the boson and fermion densities.  A key theoretical challenge  is that the pairing between bosons and fermions gives rise to \emph{fermionic} composite particles. Due to their fermionic nature, these particles evade conventional mean-field approaches and are thus much harder to describe than their bosonic counterparts  in Fermi mixtures, where they emerge as Cooper pairs or bosonic molecules. Moreover, the existence of such fermionic composites implies a phase diagram  that is richer in possible phase transitions compared to the simpler Fr\"ohlich model.  
Here we tackle this challenge by developing first steps towards a comprehensive functional-renormalization-group approach that allows access to the full phase diagram of Bose-Fermi mixtures in two dimension. Our approach accounts for the  bound-state physics arising from the extended Fr\"ohlich Hamiltonian and can be  systematically extended to describe the plethora of competing phases illustrated in Fig.~\ref{SchematicPDintro}. 

In order to explore this phase diagram it is crucial to start from limits that allow for a controlled understanding of the physics involved. One such limit is found at extreme population imbalance where just a single boson is immersed in a fermionic bath. This so-called Fermi polaron problem already displays rich physics that has been studied extensively in three dimensions~\cite{Chevy2006,Lobo2006,Combescot2007,Nikolic2007,Prokofev2008,Gubbels2008,Combescot2008,Punk2009,Mora2009,Combescot2009,Bruun2010,Schmidt2011,Parish2020}.  Here one finds that as the interaction between the impurity and the bath is tuned, the system undergoes a sharp transition from a polaronic to a molecular state. While in the polaron state the impurity is essentially weakly dressed by bath excitations, in  the  molecular state  the impurity binds tightly to one fermion of the surrounding environment giving rise to a state that, close to the transition \cite{Bruun2010,Schmidt2011}, is orthogonal to the polaron state.

The two-dimensional case has received  attention over the last decade~\cite{Zoellner2011,Parish2011,Schmidt2012,Bertaina2012,Parish2013,Kroiss2014,Vlietinck2014} as well. It turns out that this case is  more challenging to describe due to the increased significance of  quantum fluctuations  in reduced dimensions. While early works based on simple variational wave functions found no polaron-to-molecule transition~\cite{Zoellner2011}, later studies showed that this finding was in fact an artifact caused by the neglect of  three-body correlations.  Including these, one indeed recovers a polaron-to-molecule transition in two dimensions \cite{Parish2011,Parish2013}, a result supported by subsequent studies using a variety of Quantum Monte-Carlo (QMC) techniques~\cite{Bertaina2012,Kroiss2014, Vlietinck2014}. 

As the preceding discussion shows, there are strong constraints on any approach that aims to reliably describe strongly coupled Bose-Fermi mixtures in two dimensions even on a qualitative level. First, in order to address the strong-coupling character of the problem correctly, it must be based on the extended Fr\"ohlich model. Second, it must go beyond perturbation theory in order to  describe the formation of fermionic bound states. Third, at vanishing boson density it must correctly reproduce the quantum impurity limit, which necessitates the incorporation of three-body correlations. Fourth, in order to describe the phase diagram at finite boson density $n_B$, the approach must be able to deal with the fermionic nature of the composite particles that will experience Pauli blocking at finite density similar to  baryons in atomic nuclei.

All these requirements are met by the functional renormalization group (fRG). Based on an implementation of Wilson's renormalization group idea, the fRG has been successfully applied to the study of strongly coupled systems in a broad range of areas~\cite{Berges2002,Gies2012,Delamotte2012,Dupuis2020}, spanning from the asymptotic safety of quantum gravity \cite{Reuter1998a,Eichhorn2018a,Pawlowski2020a} to  high-energy~\cite{Pawlowski2014a,Strodthoff2017b},  statistical~\cite{Delamotte2004a,Benitez2012,Tarjus2020a} and condensed matter physics~\cite{Blaizot2005a,Ranccon2011a,Reuther2011a,Metzner2012,Rose2017a}. In addition to addressing the aforementioned constraints imposed by the two-dimensional polaron problem, the fRG technique developed in this work displays several other advantages. First, in contrast to variational approaches based on particle-hole excitation expansions, it provides a fully self-consistent approach that naturally includes high-order quantum fluctuations and  treats  polaron and molecular states on equal footing. Second, compared to conventional quantum field theory approaches the fRG includes quantum fluctuations in a coarse-grained fashion ---momentum-by-momentum shell--- that makes it ideally suited to treat competing ordering instabilities. Third, similar to variational techniques, the fRG  can be improved systematically by using increasingly refined truncations of the underlying quantum effective action. Finally, it offers an easier access to spectral  and dynamical response functions compared to Monte Carlo approaches where the analytic continuation of noisy data is required.

We demonstrate the applicability of our approach by focusing on the case where the size $a_B$ of the fermionic bound state  is small compared to the average distance $d\sim n_B^{-1/2}$ between bosons. Since for sufficiently short-ranged attraction this bound state always exists in two dimensions \cite{Adhikari1986}, its binding energy  $\epsilon_B=\hbar^2/(2\mu a_B^2)$ (with $\mu$  the reduced mass) is the relevant interaction scale, i.e. we work in the limit  $(\hbar^2/2\mu) n_B/\epsilon_B\ll 1$.

By including the full feedback of three-body correlations  on the renormalization group flow, we demonstrate the correct description of the polaron-to-molecule transition in the single boson limit. In particular we predict the transition to occur at a critical dimensionless interaction strength $(\epsilon_F/\epsilon_B)^\ast=1/18.78$  in excellent agreement with state-of-the-art variational \cite{Parish2011,Parish2013} and diagrammatic MC approaches \cite{Bertaina2012,Kroiss2014,Vlietinck2014}.

Having thus established the limiting case of the phase diagram, we extend  the renormalization group (RG) flow to finite boson density.  At small dimensionless Fermi energies  $\epsilon_F/\epsilon_B$, we find that
fermionic composites build up a well-defined Fermi surface, leading to the formation of a trion liquid in the case of semiconductors and a Fermi sea of dressed Feshbach molecules in the case of ultracold atoms. As the boson density is increased, the effective energy gap of the  composites decreases, leading to a transition into a strongly correlated phase where fermions are hybridized with molecular degrees of freedom. This extension of a single boson framework does not take into account the formation of higher-order bound states including more than one boson \cite{Pricoupenko2010,Levinsen2014c,Naidon2017}. While this description might thus be missing some of the phases and states at play, recent theoretical and experimental results suggest that this simplified treatment may, however, still be sufficient to describe the physics relevant on experimental times scales \cite{Guidini2015,Duda2021}.

Adapting the fRG approach to account for the full frequency-dependence of  self-energies, we predict the spectral properties of the model. We find that the inclusion of  three-body correlations has a strong impact on the effective masses of polarons and molecules (trions) which can   be  observed using state-of-the-art experimental techniques recently developed in ultracold atoms \cite{Shkedrov2018,Shkedrov2020,Ness2020}.

The paper is structured as follows: in \cref{sect:Model} we introduce the strong-coupling model of  Bose-Fermi mixtures and discuss the effective action formalism. Here we also introduce our fRG approach, derive the corresponding renormalization group  equations and discuss how the various phases discussed in this work can be distinguished. As this section contains a detailed discussion of the used technique, readers mainly interested in the predictions of our work may proceed from the introduction to \cref{sect:polaron} and the sections thereafter. In \cref{sect:polaron} we discuss the universal connection between strongly coupled Bose-Fermi mixtures in atomically thin semiconductors and ultracold atoms. We benchmark our approach on the limiting case of a  single boson embedded in a fermionic environment, obtain the ground-state energy of the system and  study the evolution of correlation functions in dependence on the fermion density and interaction strength. In \cref{sect:finitedens} we turn to the case of finite boson density. We determine the phase diagram both as a function of the chemical potential and density of both species. In \cref{sect:Kamikado} we adapt the fRG scheme to describe the spectral functions of the model and we predict the  properties of quasiparticles emerging in the theory. We conclude in \cref{sect:conclusion}, discuss perspectives for possible experimental realizations and provide an overview of open questions and promising extensions of the fRG approach introduced in the present work.

\section{Model}
\label{sect:Model}

We consider a two-dimensional Bose-Fermi mixture consisting of a fermionic  species $\psi$ into which bosonic particles $\phi$ are embedded. The system is described by the microscopic action 
\begin{align}
S &= \int_{x} \psi_x^\ast\left(\partial_\tau -\frac{\nablav^2}{2m_F} -\mu_\psi\right)  \psi^{\phantom{\ast}}_x\nnl 
&+ \int_x \phi_x^\ast\left(\partial_\tau - \frac{\nablav^2}{2m_B} -\mu_\phi\right)  \phi^{\phantom{\ast}}_x \nnl
&+g  \int_{x} \psi_x^\ast \phi_x^\ast\phi_x^{\phantom{\ast}} \psi_x^{\phantom{\ast}} \label{singlechannel}
\end{align}
where $x=(\rv,\tau)$ denotes the coordinate $\rv$ and imaginary time  $\tau \in [0,1/T]$; moreover, $\int_x=\int_0^{1/T} d \tau \int d^2 \rv$. In the following, we consider zero temperature, $T=0$,  and  assume that  bosons and fermions  have the same mass $m=m_F=m_B$. We work in units $\hbar=k_\mathrm{B}=1$, and set $2m=1$ unless indicated otherwise. The fields $\psi$ and $\phi$ are of fermionic Grassmann and complex boson nature, respectively. The two species interact by means of an attractive contact potential of strength $g<0$. The model is regularized in the ultraviolet (UV) by a momentum cutoff $\Lambda$. 

The densities of both species are set by the chemical potentials $\mu_{\psi/\phi}$ \footnote{For a non-interacting system, i.e. when $g=0$, the chemical potential $\mu_\phi \equiv 0$ at $T=0$. In this case, the density of the bosons is set by the choice of the state of the system, e.g. in the form of a boson coherent state.}. At a finite fermion density $n_\psi$ (set by a chemical potential $\mu_\psi>0$), tuning the boson chemical potential $\mu_\phi$ at fixed $\mu_\psi$ and $g$,  triggers a transition at a critical chemical potential $\mu_\phi^c$  between a vacuum phase of bosons ($\mu_\phi<\mu_\phi^c$) with vanishing boson density to a phase of finite boson density  $n_B>0$ ($\mu_\phi>\mu_\phi^c$).

For strongly coupled Bose-Fermi mixtures it is crucial to allow for the possibility of the pairing of the bosons and fermions to a composite fermionic molecular (trion) state. In order to describe this bound state, it is essential to resolve the pole structure of the scattering vertex sufficiently well \cite{Combescot2006,Combescot2007}. In order to achieve this  in an efficient way, rather than considering the action in \cref{singlechannel}, we study a two-channel model where the interspecies interaction is mediated by a molecule field $t$, that describes a composite fermionic particle of mass $2m$  \cite{Holland2001,Timmermans2001,Bruun2004,Bloch2008}. The action is given by
\begin{align}
S &= \int_{\pv,\omega}\bigg\{ \psi_P^\ast\left(-i\omega +\pv^2 -\mu_\psi\right)  \psi_P\nnl 
&+ \phi_P^\ast\left(-i\omega +\pv^2 -\mu_\phi\right)  \phi^{\phantom{\ast}}_P +t^{\ast}_P\left(-i\omega + \dfrac{\pv^2}{2} + m_{t}\right) t^{\phantom{\ast}}_P \bigg\} \nnl
&+h\int_{x}\left\{ \psi^{\ast}_x \phi^{\ast}_x t^{\phantom{\ast}}_{x}+t^{\ast}_{x}\phi^{\phantom{\ast}}_x\psi^{\phantom{\ast}}_x\right\}\ . \label{twochannel}
\end{align}
Here, a boson and a fermion can be converted into the molecule (trion) $t$ with a conversion Yukawa coupling $h$, and $m_{t}$ is the detuning energy of the molecule. In Eq.~\eqref{twochannel} we give the action in Fourier space where  $P=(\pv,\omega)$ comprises the momentum $\pv$ and the Matsubara frequency $\omega$, and  $\int_{\pv,\omega}\equiv \int d^2 \pv d\omega$. 
 We operate in  the limit where $h\to\infty$ which universally describes  both open-channel dominated Feshbach resonances in cold atoms \cite{Chin2010} as well as electron-exciton scattering in atomically thin transition-metal dichalcogenides \cite{Fey2020}. In this limit,  $t$ becomes a purely auxiliary Hubbard-Stratonovich field, i.e. it can be integrated out to yield  back  the original action \labelcref{singlechannel} when $h^2/m_{t}=-g$ is fulfilled~\cite{Lurie1964,Nikolic2007}.
 
In two dimensions,  a bound state exists for any attractive interaction strength $g<0$ \cite{Adhikari1986}. Using a sharp UV cutoff in the Lippmann-Schwinger equation, the binding energy $\epsilon_B$ is related to the parameters of the microscopic model through~\cite{Adhikari1986,Randeria1990,Zoellner2011} 
\begin{equation}
m_{t  }= \dfrac{h^2}{8\pi} \log\bigg(1 + \dfrac{2\Lambda^2}{\epsilon_B} \bigg)\ . \label{mphiinitial}
\end{equation}  
Thus, rather than using the microscopic coupling $g$ (or equivalently $h$ and $m_t$) we can parametrize the interaction strength in terms of the experimentally measurable binding energy $\epsilon_B$ of the molecule (in the case of cold atoms) or trion (in the case of 2D semiconductors), respectively. Note, in the following we will use the terms trion and molecule often interchangeably.
 
\subsection{fRG formalism and effective action}
The fRG is a momentum space implementation of Wilson's renormalization group. In the following, we briefly recall its principle; for a detailed discussion we refer to Refs.~\cite{Berges2002, Delamotte2012,Metzner2012,Gies2012}. The idea behind the fRG is to build a family of theories indexed by a momentum scale $k$ such that only quantum fluctuations above that scale are taken into account. Thus rather than treating fluctuations at all scales at once, one iteratively integrates out  modes from high to low energies by smoothly lowering $k$ from the microscopic UV scale $\Lambda$ down to $k=0$.

In practice this is done by adding to the action \labelcref{twochannel} an infrared regulator term
\begin{align}
\Delta S_k =& \int_{\pv,\omega} \bigg\{ \psi_{P}^\ast R^{\phantom{\ast}}_{\psi,k}(P)\psi^{\phantom{\ast}}_{P}+ \phi_{P}^\ast R^{\phantom{\ast}}_{\phi,k}(P)\phi^{\phantom{\ast}}_{P}\nnl
+& t^\ast_{P} R^{\phantom{\ast}}_{t,k}(P)t^{\phantom{\ast}}_{P} \bigg\} 
\label{eq:regu0}
\end{align}
which penalizes low-energy fluctuations, such that only high-energy modes contribute to the field integral.

For bosons and  fermions at vanishing density the low-energy modes are located at small momenta. Thus the cutoff function $R_{\sigma,k}(P)$ ($\sigma=\psi$, $\phi$, $t$) is set to be large (wrt. $k^2$) for $|\pv|\ll k$ and negligible  for $|\pv| \gg k$. In this way, low-momentum fluctuations are suppressed while high-momentum ones are left unaffected. For fermions at a finite density,  the low-energy modes are located around the Fermi surface. Accordingly, in this case $R_{\sigma,k}$ is chosen to suppress fluctuations of modes inside a momentum shell of width $\sim 2k$ around the Fermi surface.

Starting from the sum of $S$ and $\Delta S_k$ one then defines a scale-dependent partition function $\Zc_k$, as well as a scale-dependent effective action $\Gamma_k$ through a (modified) Legendre transform of the free energy $\ln \Zc_k$. The evolution or `flow' of the effective action as the scale $k$ is lowered is then given by the Wetterich equation \cite{Wetterich1993},
\begin{align}
\partial_k \Gamma_k = \frac{1}{2} \STr \Big[ \Big(\Gamma_k^{(2)}+ R_k\Big)^{-1}\partial_k R_k \Big]\ . \label{wetterich}
\end{align}
In the above expression, the supertrace $\STr$ denotes a summation over all momenta and frequencies, as well as  the different fields, including a minus sign for fermions. Moreover, $\Gamma_k^{(2)}$ and $R_k$ represent the matrices of second functional derivatives of $\Gamma_k$ and $\Delta S_k$, respectively, with respect to the quantum fields \footnote{The quantum fields are defined as the expectation value of the particle fields in presence of their source, i.e. $\psi[J_\psi] \equiv \frac{\delta \ln \Zc_k[J_\psi]}{\delta J^*_{\psi}}$. The external source $J_{\psi}$ is a Grassmann or complex field that couples linearly to the field $\psi$ within $\Zc_k$.  Furthermore, we use the same symbol for the fields in the action $S$ and in the flowing effective action $\Gamma_{k}$.}.

Provided $R_{k=\Lambda}=\infty$ \footnote{In practice $R_{k=\Lambda}\simeq\Lambda^2$ is sufficient.} at $k=\Lambda$ all fluctuations are suppressed and $\Gamma_{k=\Lambda}=S+const.$ \cite{Delamotte2012,Gies2012}. On the other hand, for $R_{k=0}=0$ one recovers at $k=0$  the effective action of the original model, $\Gamma_{k=0}=\Gamma$. Crucially, the effective action $\Gamma$ (Gibbs free energy) is the generating functional of all one-particle irreducible vertices. It thus  contains all  information about the exact solution of the theory and hence its determination corresponds to solving the non-relativistic, many-body Schr\"odinger equation.

\subsection{Truncation schemes}
\label{sec:truncation}

While the flow equation \eqref{wetterich} is exact, it is, in most practical cases, impossible to solve without resorting to approximations. A standard strategy is to propose an \emph{Ansatz} for  the flowing effective action $\Gamma_k$. When dealing with fermions, it is customary to expand in the powers of the fields in a so-called vertex expansion~\cite{Metzner2012}. Following this strategy we choose the \emph{Ansatz} for the field-dependent part of the effective action
\begin{align}
\Gamma_{2,k}= {}&\int_{\pv,\omega} \bigg\{ \psi_P^\ast G_{\psi,k}^{-1}(P)  \psi_P^{\phantom{\ast}}+\phi_P^{\ast}G^{-1}_{\phi, k}(P) \phi_P^{\phantom{\ast}}\nonumber \\
&{}+ t_P^\ast G_{t,k}^{-1}(P)  t_P^{\phantom{\ast}} \bigg\}
+h_k\int_{x}  (\psi^{\ast}_{x} \phi^{\ast}_{x} t_{x}^{\phantom{\ast}}+t^{\ast}_{x}\phi_{x}^{\phantom{\ast}} \psi_{x}^{\phantom{\ast}})\ . \label{truncation2}
\end{align}

We perform an additional gradient expansion by neglecting a possibly emerging momentum dependence of the Yukawa coupling $h_k$ via vertex corrections. Each field $\sigma=\psi,\phi,t$ carries  renormalized flowing single-particle Green's functions whose momentum dependence is approximated within the gradient expansion as 
\begin{align} 
 G_{\psi,k}^{-1}(\pv,\omega)={}&A_{\psi,k}\left(-i\omega +\pv^2 - \mu_{\psi} \right) +m_{\psi,k}\ , \label{eq:AnsatzPropFer}\\
 G_{\phi,k}^{-1}(\pv,\omega)={}&A_{\phi,k}\left(-i\omega +\pv^2 \right) +m_{\phi,k}\ , \label{eq:AnsatzPropBos}\\
 G^{-1}_{t, k}(\pv,\omega)={}&A_{t,k}  \left(-i\omega + \pv^2 /2 \right)+ m_{t,k}\ , \label{eq:AnsatzPropMol}
\end{align}
 parametrized by inverse quasiparticle weights $A_{\sigma,k}$ and detunings $m_{\sigma,k}$. Note that for the boson field $\phi$ we have absorbed the dependence on the chemical potential $\mu_\phi$ into the definition of the detuning $m_{\phi,k}$ for convenience. The \emph{Ansatz} \eqref{truncation2} incorporates in detail two-body correlations between the bosons and fermions. In particular,  it describes well the pairing correlations between the particles which is essential to enter the strong-coupling regime. As a short-hand we refer to the effective flowing action \eqref{truncation2} as the \textit{`two-body truncation'}.

The two-body truncation has been used  successfully to study the Fermi polaron problem in three space dimensions~\cite{Schmidt2011, Kamikado2017,Pawlowski2017}. In two space dimensions, however, quantum fluctuations are stronger and previous works~\cite{Parish2011,Zoellner2011} have established  that higher-order correlations  must be taken into account to  describe the ground state of the system. Indeed, as we shall see in \cref{sect:polaron}, the two-body truncation is not sufficient to describe the polaron--to--molecule transition.

\begin{figure}[t]
	\begin{center}
		\includegraphics[width=1\linewidth,origin=c]{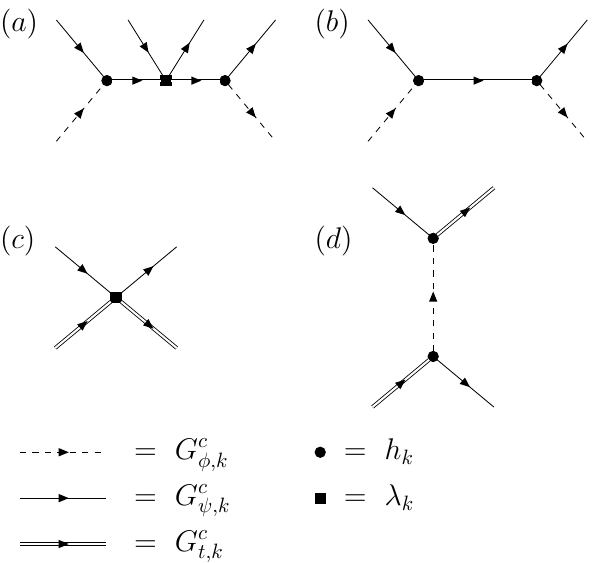}
	\end{center}
	\caption{(a) Tree-level diagram generated from the effective action that leads to the emergence of three-body correlations in the system. (b) Exchange tree-level diagram giving rise to the Bose-Fermi scattering $T$-matrix. (c), (d) Tree-level diagrams contributing to the overall atom-molecule scattering amplitude. The dashed, solid and double lines denote the boson, fermion and molecule Green's functions $G_{\phi/\psi/t,k=0}$, respectively, while the dots and squares denote $h_{k=0}$ and $\lambda_{k=0}$, all evaluated at $k=0$.
	} \label{fig:treelevelthreebody}
	\end{figure}

Consequently, we extend the \emph{Ansatz} for the effective action to a \textit{`three-body truncation'}. To this end we add a term to the two-body truncation that accounts for the build up of three-body correlations during the RG flow:
\begin{align}
\Gamma_{3,k} = {}&\Gamma_{2,k} +\lambda_k \int_{x}  \psi^{\ast}_x t_{x}^{\ast}t^{\phantom{\ast}}_{x} \psi^{\phantom{\ast}}_{x} \ . \label{truncation3}
\end{align}
The additional term proportional to the contact coupling $\lambda_{k}$ describes the scattering between composite molecules and  fermions, and thus, by virtue of the tree-level diagram depicted in Fig.~\ref{fig:treelevelthreebody}(a), it accounts effectively for the emergence of three-body correlations in the system.

Let us  briefly comment on the validity of the gradient expansion used for both truncations \eqref{truncation2} and \eqref{truncation3}. In the single-boson limit, we expect the low-energy excitations of  the boson $\phi$ and the composite particle $t$ to be at small momenta and we  may thus expand the momentum-dependence of their propagators in a power series about $\pv=0$, $\omega=0$. For the fermions,  on the other hand, we expect the most relevant excitations to lie around the Fermi surface. We thus expand their propagator about $\pv^2=\epsilon_F$, $\omega=0$. 

As we extend our calculation to a finite boson density we retain the expansion around $\pv=0$ for the molecules as we will find  that their phase appears in a regime of the phase diagram where  $n_B\ll n_F$. Thus the Fermi energy of molecules always remains small. Moreover, we employ a gradient expansion that neglects effective mass corrections as these are not expected to be crucial to correctly capture the qualitative physics of the phase diagram (except for large mass ratios $m_F\ll m_B$ \cite{Levinsen2013}, a regime not considered in this work).

In the quantum impurity limit, the vanishing of the boson density implies that the properties of the fermionic Green's function are not affected by interactions; i.e. the propagator in Eq.~\eqref{eq:AnsatzPropFer} with $A_{\psi,k}=1$ and $m_{\psi,k}=0$ is \emph{exact}. This can also be verified explicitly  from the flow equations derived further  below [cf. \cref{Gphiflow,Gpsiflow,Gtflow,hflow,lambdaflow}]. At finite boson density we neglect the renormalization of the fermionic  propagators since throughout this work we will remain in the regime of  density ratios $n_B\ll n_F$. 

While the truncation in Eq.~\eqref{truncation3} can be improved systematically, e.g., by considering higher-order correlations or a more involved momentum dependence of the propagators or the vertices, the model in Eq.~\eqref{truncation3} is sufficient to accurately describe the intricate quantum impurity limit, as shown in \cref{sect:polaron}.  In particular, even though $h_k$ and $\lambda_k$ have no momentum dependence,  the  Bose-Fermi scattering $T$-matrix, as described by the exchange tree-level diagram shown in Fig.~\ref{fig:treelevelthreebody}(b), acquires a momentum dependence due to the dynamic field $t$ that is sufficient to describe accurately the Bose-Fermi scattering at the relevant energy scales. 

We note that at finite boson density our truncation does not account for the possible formation of bound states between two or more bosons and a single fermion~\cite{Pricoupenko2010,Naidon2017,Levinsen2014c}. In a realistic  experimental setting, where the system will be prepared adiabatically, the formation of these higher-order bound states requires several bosons to be located in the close vicinity of the fermions. Since we focus here, however, on the regime where the boson density is significantly smaller than the fermion density, $n_B\ll n_F$, the probability to find such configurations will be small. As a result, compared to the time scale of Fermi polaron or molecule formation, the formation of higher-order bound states will be suppressed, enabling the observation of the phase diagram studied in this work on transient time scales. Nevertheless, while recent results suggest that this treatment is appropriate \cite{Guidini2015,Duda2021}, the framework used in this work can be extended to feature bound states between two bosons and a fermion; both in the vacuum limit as well as at finite density (for details see \cref{BBFcoupling}). This highlights that this study provides only an initial step in the exploration of this phase diagram which, given sufficiently stable bound states, may feature an even richer structure.

\subsection{Regulators}\label{sec:reg}
For the regulators $R_{\sigma,k}$ we use sharp cutoff functions ~\cite{Metzner2012}, defined so that the \emph{regulated} inverse flowing propagators
\begin{equation}
(G_{\sigma,k}^c)^{-1} = (G_{\sigma,k})^{-1} + R_{\sigma,k}
\end{equation} 
appearing on the  rhs. of the flow equation \labelcref{wetterich} acquire the simple form
\begin{align}
G^{c}_{\psi,k}(\pv,\omega)&= G_{\psi,k}(\pv,\omega)\Theta(|\pv^2 -\epsilon_{F,k}|-k^2) \ ,  \label{regulatorPsi}\\
G^{c}_{\phi,k}(\pv,\omega)&=G_{\phi,k}(\pv,\omega) \Theta(|\pv|-k) \ ,  \label{regulatorPhi}\\
G^{c}_{t,k}(\pv,\omega)&= G_{t,k}(\pv,\omega)\Theta(|\pv|-k) \ . \label{regulatorT}
\end{align}
Here,  $\epsilon_{F,k}= \mu_\psi- m_{\psi,k}/ A_{\psi,k}$ is the Fermi energy of the fermionic species $\psi$. For the fermions $\psi$ the regulator suppresses  fluctuations at momenta in a shell of width $2k$ around the, in principle, flowing Fermi-surface of the bath \cite{Floerchinger2010}. Even though the molecule is a fermion as well and thus may develop a Fermi surface at finite boson density, we regulate it about zero momentum as all phases considered in this work appear in the regime $n_B \ll n_F$.

The choice of sharp cutoff functions has several advantages \footnote{For a detailed discussion of regulator dependence in this model in three  dimensions see Ref.~\cite{Pawlowski2017}.}. Foremost, it allows for an analytic derivation of the flow equations. In addition, it facilitates the comparison to  previous FRG studies~\cite{Schmidt2011,Kamikado2017,Pawlowski2017} as well as to self-consistent diagrammatic  approximations that display a similar mathematical structure~\cite{rath2013}. 

\subsection{Flow Equations}\label{sec:flowequations}

\begin{figure}[t]
	\begin{center}
		\includegraphics[width=\linewidth]{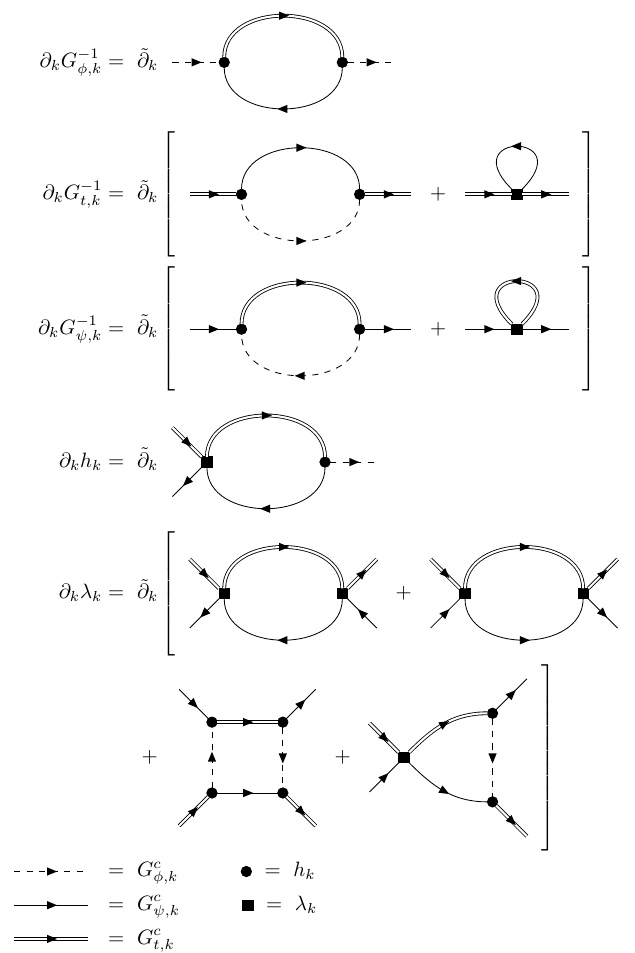} 
	\end{center}
	\caption{Diagrammatic representation of the fRG flow equations \labelcref{Gphiflow,Gpsiflow,Gtflow,hflow,lambdaflow}. Lines represent the full scale-dependent propagators, including the regulators, and the dots denote interaction vertices.} \label{floweq} \end{figure}

We now turn to the explicit derivation of the RG equations \cite{Huber2012,Huber2020} of all running coupling constants. For the three-body truncation $\Gamma_{3,k}$  ($\Gamma_{2,k}$ is a subset obtained by setting $\lambda_{k}\equiv 0$ in all flow equations) all vertices can be expressed in terms of the six running couplings $A_{\sigma,k}$, $m_{\sigma,k}$, $h_k$ and $\lambda_k$. Following the prescription detailed in \cref{sect:AppProj}, the flow equations are obtained from appropriate functional  derivatives of the Wetterich equation. Their diagrammatic representation is shown in \cref{floweq} and in terms of the flowing Green's functions they read
\begin{align}
\partial_k G^{-1}_{\phi,k}(P)&= h_k^2 \tilde{\partial}_k \int_Q G^c_{t,k}(P+Q) G^c_{\psi,k}(Q) \ ,\label{Gphiflow}\\
\partial_k G^{-1}_{\psi,k}(P)&= - h_k^2 \tilde{\partial}_k \int_Q\Big[ G^c_{t,k}(P+Q) G^c_{\phi,k}(Q)\nnl
&\ \ \ \ \ \ \ \ \ \ \ \ \ \  \ \ \  +\frac{\lambda_k}{h_k^2 }G^c_{t,k}(Q)\Big] \ , \label{Gpsiflow}\\
\partial_k G^{-1}_{t,k}(P)&=- h_k^2 \tilde{\partial}_k \int_Q \Big[G^{c}_{\phi,k}(P-Q) G^{c}_{\psi,k}(Q)\nnl
&\ \ \ \ \ \ \ \ \ \ \ \ \ \  \ \ \ +\frac{\lambda}{h_k^2} G^c_{\psi,k}(Q)\Big] \ ,\label{Gtflow}
\end{align}
and
\begin{align}
\partial_k h_k &=-\frac{\lambda_k}{h_k} \partial_k G^{-1}_{\phi,k}(0) \ ,  \label{hflow}\\
\partial_k \lambda_k &=  -\lambda_k^2\tilde{\partial}_k \int_Q G^c_{t,k}(Q)\left[G^c_{\psi,k}(Q)+ G^c_{\psi,k}(-Q)\right]\nonumber\\
&-h_k^4\tilde{\partial}_k \int_Q G^c_{t,k}(Q)G^c_{\phi,k}(Q)^2 G^c_{\psi,k}(-Q)\nonumber\\
&-2 h_k^2 \lambda_k \tilde{\partial}_k \int_Q G^c_{t,k}(Q)G^c_{\phi,k}(Q) G^c_{\psi,k}(-Q)\ . \label{lambdaflow}
\end{align}
In these expressions $\tilde{\partial}_k$ stands for the derivative with respect to the $k$ dependence of the regulator only, i.e. $\tilde{\partial}_k=(\partial_k R_k) \partial_{R_k}$ and $\int_P\equiv(2\pi)^{-3}\int d^2 \pv d\omega$. As discussed in \cref{sect:gradexppar}, from \cref{Gphiflow,Gpsiflow,Gtflow} the flow equations of the couplings $A_{\sigma,k}$, $m_{\sigma,k}$ are obtained by projection onto the momentum dependencies given in \cref{eq:AnsatzPropBos,eq:AnsatzPropFer,eq:AnsatzPropMol}.

\subsection{RG initial conditions}
\label{sec:RGInitial}

The initial conditions for the flows are obtained by setting $\Gamma_{k=\Lambda}=S +const$. First, we discuss the UV initial conditions for $m_t$ and $A_t$ which are obtained as follows. The two-body problem of a single boson scattering with a single fermion can be solved exactly. The resulting initial condition for $m_{t,k=\Lambda}$ is given by \cref{mphiinitial}. To arrive at this expression one may recognize that in the two-body problem the molecule is the ground state. As such it has to be a gapless degree of freedom in the infrared, i.e. $m_{t,k=0}=0$.
Moreover,  $\mu_{\psi}$ and $\mu_{\phi}$ must be set to negative values $\mu_{\psi}^{\text{vac}}$ and $\mu_{\phi}^{\text{vac}}$ to yield a vanishing density of either species. In addition,   $\mu_{\psi}^{\text{vac}}+\mu_{\phi}^{\text{vac}}=-\epsilon_B$ has to be  fulfilled to ensure that the energy cost to create two  particles from the vacuum to form a bound state is given by the molecular binding energy $\epsilon_B$. Using these conditions, together with the fact that in the two-body problem neither the boson and fermion propagators $G_{\psi,k}$ and $G_{\phi,k}$ nor the vertices $h_k$ and $\lambda_k$ renormalize, the flow of $m_{t,k}$ can be solved analytically to yield \cref{mphiinitial} (for more details we refer to \cref{app:initial}). 

We work in the limit of large $h_{k=\Lambda}$ which ensures $t$ to be purely an auxiliary field and we use $A_{t,k=\Lambda} = 1$. Furthermore, we set $\lambda_{k=\Lambda}=0$ as it does not appear in the classical action in \cref{twochannel}.

The initial condition for the field renormalization of the boson field is naturally given by $A_{\phi,\Lambda}=1$, and the UV value of its detuning is set by the boson chemical potential, $m_{\phi,\Lambda} = -\mu_\phi$. Finally, since we will study only phases at small ratios $n_B\ll n_F$ we can assume that the fermion field is not renormalized, i.e. $A_{\psi,k} = 1$ and $m_{\psi,k}=0$ throughout the RG flow.

\subsection{Chemical potentials and distinction of phases} \label{sec:chempot}

The numerical integration of the flow equations yields the physical value of the propagators and interaction vertices at the infrared scale $k=0$. Depending on their properties we can distinguish various states and phases of the strongly coupled Bose-Fermi mixture, summarized in Table \ref{phasedefs}.

In the single-boson limit, yet at finite fermion density, we distinguish two \textit{states}: a \textit{molecular state} in which the boson is paired into a composite particle, and a \textit{polaron state} where the boson is dressed by fluctuations of majority fermion particles. At finite boson density, we distinguish two \textit{phases}: a \textit{molecular phase}, where all bosons are paired into fermionic molecules, $n_t > 0 $ and $n_\phi=0$, and a \textit{mixed phase} where molecules and unpaired polarons coexist \cite{Powell2005}. In the mixed phase,  $n_\phi>0$, so that the condensate of bosons creates a bilinear coupling in the effective action $\sim h \sqrt{n_\phi} (t^* \psi+\mathrm{c.c.})$  leading to a hybridization of the fermions with the molecular degree of freedom. This means that no purely polaronic phase with $n_t=0$ and $n_\phi>0$ is possible. In the limit of $n_B\to 0 $ the mixed phase connects to the polaron state, whereas the  molecular phase connects to the molecular state. 

In order to differentiate between these states and phases we consider the different densities defined by integrals proportional to $\int_{\pv,\omega} G_{\sigma, k=0 } (\pv, \omega )$. These densities are nonzero only when poles of $G_{\sigma, k=0 }(\pv, \omega)$ lie in the upper half of the complex $\omega$-frequency plane. Hence, from the location of poles, manifest in the energy gaps of the particle in the infrared, we can determine whether the corresponding densities vanish.

Specifically, the boson vacuum corresponds to a finite excitation gap for both the boson and the molecule, $m_{\phi,k=0}$, $m_{t,k=0}>0$. Likewise, in the single-boson limit the ground state has to be gapless while the excited state is gapped since this limit marks the boundary between the boson vacuum and the many-boson regime. 

\begin{table}[t]
\centering
	\begin{tabular}{|c |  c | c |c|c|c| c |}\hline  
	$n_{t}$&$n_\phi$& $\mu_\phi$ & $m_{t,0} $ & $ m_{\phi,0}$ & state/phase  & \# bosons\\
	\hline
	$=0$ &$=0$& $< \mu_{\phi}^c$ & $>0$ &$ >0$  & boson-vacuum &0 \\
   $=0$ &$=0$& $= \mu_{\phi}^c$ & $=0$ &$ >0$ & molecular state & 1  \\
   $=0$ &$=0$& $= \mu_{\phi}^c$ & $>0$ &$ =0$ & polaron state&1 \\
    $>0$ &$=0$&  $> \mu_{\phi}^c$ & $<0$ &$ >0$ & molecular phase &$\gg 1$ \\
  $>0$ &$>0$& $> \mu_{\phi}^c$ & $\in \mathbb{R}$  &$ =0$ & mixed phase &$\gg 1$ \\ \hline
	\end{tabular}	
 	\caption{Characterization of different ground states and phases of strongly coupled Bose-Fermi mixtures discussed in this work. }\label{phasedefs}
\end{table}

\begin{table*}[t]
  \centering
  {\renewcommand{\arraystretch}{1.2}
  \begin{tabular}{|lr|c|c|}
  \hline 
    
     && 2D semiconductors (TMD) & cold atoms \\ \hline
     
    fermions& & electron/hole & atom \\
    &charge   & negative/positive & neutral \\ 
    &size & pointlike & $\sim a_0$ \\ \hline
        bosons& & exciton & atom \\
        &charge & neutral & neutral\\ 
        &size & $\sim 1 \text{nm}$ & $\sim a_0$ \\ \hline
    composite fermion && trion & molecule \\
    &charge & charged & neutral \\ 
    &size & $\sim 2\text{nm}$ $(\text{fixed})$ & $\sim 1000 a_0 $ (tunable)\\ \hline 
    typical Fermi energy $\sim \epsilon_F$ && $0-50 \text{meV} \ \widehat{=}\  \text{0 - 10THz}$& $50 \text{peV} \ \widehat{=}\  10 \text{kHz}$ \\ 
    &tunability & tunable: gate doping & $\sim$fixed \\ \hline 
    typical interaction energy $\sim\epsilon_B$ && $30 \text{meV} \ \widehat{=}\  \text{10THz}$ & $0-50 \text{peV}\ \widehat{=}\ 0-10 \text{kHz}$\\
  
 &tunability & fixed & tunable: Feshbach resonances\\ \hline
 dimension && 2D & 1D, 2D, 3D \\ \hline 
 Fermi temperature  $T/T_F$ && $\text{mK}-300\text{K}$: $T/T_F\sim 0-2$& $5\text{nK}-\mu \text{K}$: $T/T_F\sim0.05 - 2 $  \\ \hline
boson-fermion potential && short-ranged, polarization int. $\sim 1/ r^4$ & short-ranged, vdW/Feshbach int.  \\ \hline
inter-fermion separation && $>1\text{nm}$ (tunable) &   $\sim 1000 a_0 $ $(\sim\text{fixed})$ \\ \hline

dimensionless interaction strength $\epsilon_B/\epsilon_F$ && $\sim 1$, strong coupling & $\sim1$, strong coupling \\ \hline 
 \end{tabular}}
  \caption{Comparison of key properties of physical systems in which two-dimensional Bose-Fermi mixtures can be realized in a universal way: two-dimensional semiconductors hosted in atomically thin transition-metal dichalcogenides (TMD) and confined, quasi-two-dimensional  gases of ultracold atoms interacting via Feshbach resonances. The constant $a_0=0.529  $\r{A}  denotes the Bohr radius.}
  \label{tab:1}
\end{table*}

At finite boson density, the molecular phase corresponds to $m_{\phi,k=0}>0$ and $m_{t,k=0}<0$, i.e.  molecules feature a Fermi surface determined by their Fermi energy $-m_{t,k=0}/A_{t,k=0}$. For the mixed phase, the situation is more subtle. Our \emph{Ansatz} does not allow for the description of a condensate at  finite boson density that could be accounted for, e.g., by shifting the $\phi$-field expectation value by a coherent state transformation. However, it is still possible to predict whether a boson condensate forms. Indeed, a necessary condition for the existence of a $\phi$-condensate is that for some $0\leq k \leq \Lambda$ the boson gap $m_{\phi,k}$ vanishes \footnote{When the condensate appears at finite RG scale $k>0$ it could, of course, again vanish at smaller RG scales due to the effect of quantum or thermal fluctuations.}. In that case, even though we are unable to further pursue the RG flow, we identify the phase to be the mixed phase.

In this mixed phase the bilinear term mentioned above leads to a mixing of the fermionic and the molecular propagators. Consequently, these propagators share the same pole structure and the corresponding species are thus populated simultaneously. As a result all three particle species are present in this phase. This implies that in our model a  regime   populated exclusively by majority fermions and condensed minority bosons is   possible only  in the single-boson limit at $n_B=0$.

\section{Quantum Impurity Limit: single boson in a Fermi sea}
\label{sect:polaron}

We first apply our approach to the limiting case of the Bose-Fermi phase diagram where an individual boson is immersed in a  bath of fermions. This limit defines the so-called Fermi polaron problem, and its solution determines the phase diagram along the $y$-axis of Fig.~\ref{SchematicPDintro}. In order to reach this single-boson limit, the boson chemical potential is tuned to the critical value $\mu_\phi = \mu_{\phi}^c$ that separates the boson vacuum ($\mu_\phi<\mu_{\phi}^c$) from the phase of a finite boson density ($\mu_\phi>\mu_{\phi}^c$); see Table \ref{phasedefs}.

\subsection{Fermi polaron problem in ultracold atoms and atomically thin semiconductors}

 The nature of the ground state of the Fermi polaron problem universally depends on the ratio of the two relevant energy scales of the problem: the kinetic energy, represented by $\epsilon_F$, and the interaction energy, set by $\epsilon_B$. While $\epsilon_F/\epsilon_B$ can, in theory,  be tuned by adjusting either  $\epsilon_F$ or $\epsilon_B$, in experiments it depends on the physical system which parameter is accessible for easy tunability.
 
 The two main systems in which strongly coupled Bose-Fermi mixtures can be realized today are ultracold atoms and atomically thin semiconducting transition-metal dichalcogenides (TMD). To support the following discussion, in \cref{tab:1} we summarize key parameters and quantities describing the universal connection between these systems.

 In monolayer TMD, $\epsilon_B$ represents the trion binding energy which is typically fixed \cite{Sidler2016,Courtade2017,Fey2020,Imamoglu2021,Raja2017}. However, by electrostatically doping the system with charge carriers, the Fermi energy $\epsilon_F$ is easily adjusted and $\epsilon_F/\epsilon_B$ can thus be tuned. In cold atoms the situation is reversed. Here, the binding energy $\epsilon_B$  can be tuned using Feshbach resonances, while adjusting the Fermi energy over a wide range of values is challenging.  As a result, in cold atoms the Fermi energy $\epsilon_F$ is the natural unit and, correspondingly,  the spectrum of the system is  expressed as a function of the dimensionless energy $E/\epsilon_F$ and interaction strength $\epsilon_B/\epsilon_F$. In contrast, in TMD the binding energy $\epsilon_B$ provides the appropriate unit, and the spectrum is expressed as a function of $E/\epsilon_B$ and  $\epsilon_F/\epsilon_B$. 

Of course, physics does  not depend on the chosen units. It is, however, still instructive to compare spectra for both sets of units, as the choice of units reflects the experimental protocols employed to observe the physics of Fermi polarons: in TMD  using  gate-doping of $\epsilon_F$ and in cold atoms interaction tuning of $\epsilon_B$ exploiting Feshbach resonances.

\subsection{Quasiparticle energies}

\begin{figure}[t]
	\begin{center}
    \includegraphics[width=\linewidth]{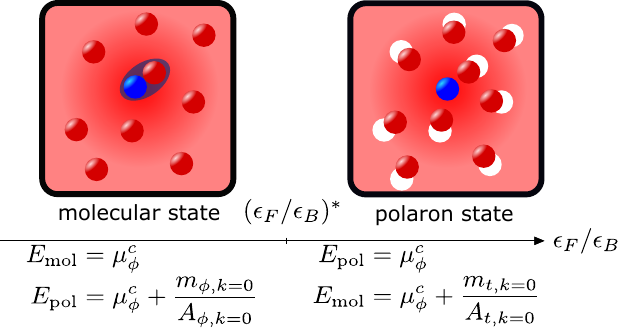}
	\end{center}
	\caption{Schematic characterization of the relevant states in the  Fermi polaron problem. In the molecular state the impurity binds to a single fermion while in the polaron state it is collectively dressed by the environment. For each state, we give the expressions for the energies  $E_{\text{pol}}$ and $E_{\text{mol}}$ of the polaron and the molecule.}\label{fig:schematicSinglePD} \end{figure}

In order to obtain the spectrum of the Fermi polaron problem we first determine the ground-state energy of the system, set by $\mu_\phi^c$, the critical energy needed to bring a boson from the vacuum.
The procedure is summarized in Fig.~\ref{fig:schematicSinglePD}: when the polaron  is the ground state, $m_{\phi,k=0}=0$, and the polaron energy is given by $E_\text{pol}=\mu_\phi^c$. In this \textit{`polaron regime'} the molecular state is an excited state whose energy  is determined from the pole of its Green's function which yields $E_\text{mol}=E_\text{pol}+m_{t,k=0}/A_{t,k=0} $. In turn, in the \textit{`molecular regime'} the molecule is the ground state. Here, $m_{t,k=0}=0$, and the molecule energy is given by $E_\text{mol}= \mu_\phi^c$, while the polaron is an excited state with an energy gap $E_\text{pol}=E_\text{mol}+m_{\phi,k=0}/A_{\phi,k=0}$.

\begin{figure*}[t]
	\begin{center}
			\includegraphics[width=\linewidth]{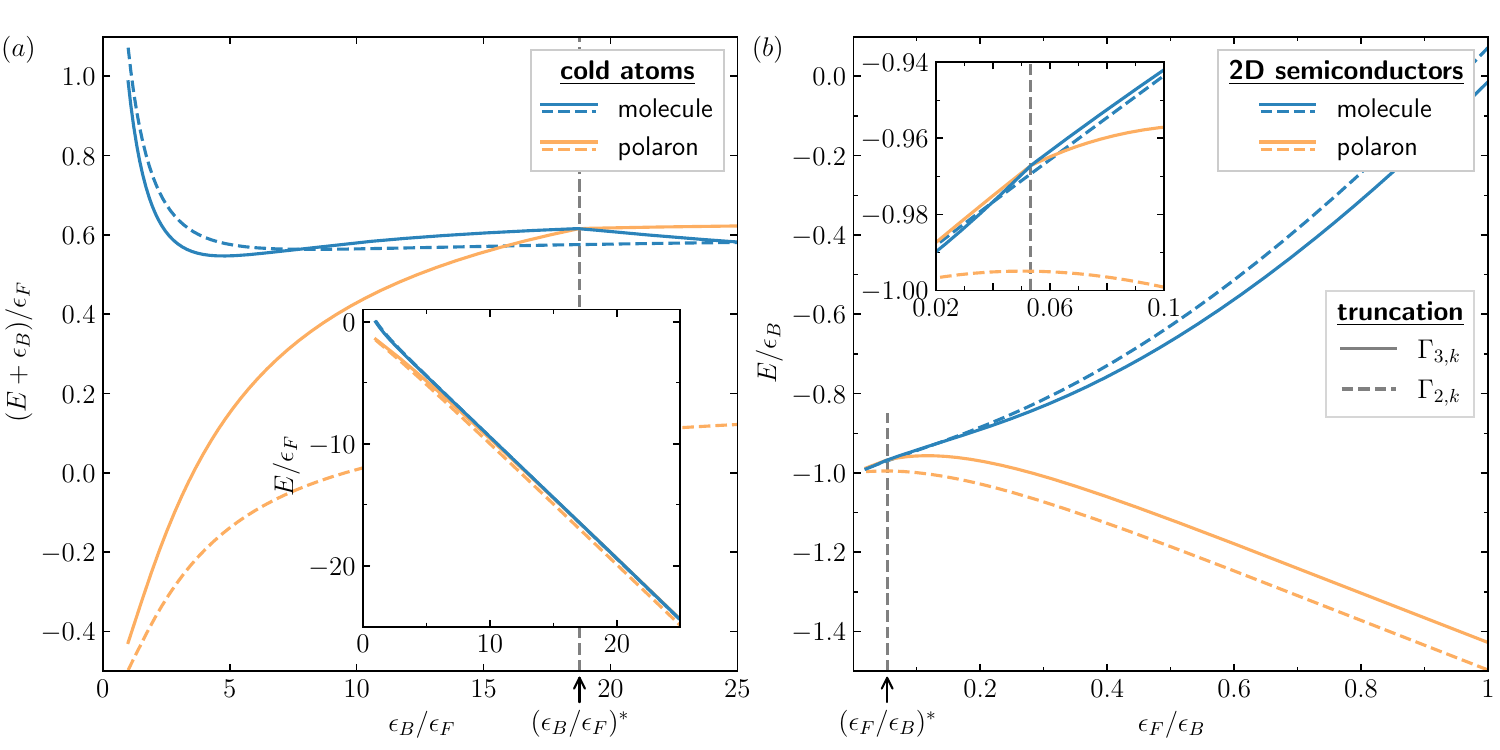}
	\end{center}
	\caption{Polaron and molecule energies, $E_\text{pol}$ (orange)  and $E_\text{mol}$ (blue), obtained using the two- and three-body truncations  $\Gamma_{2,k}$ (dashed lines) and $\Gamma_{3,k}$ (solid lines). (a) Energy spectrum expressed in units most suitable for cold atom experiments where $\epsilon_B$ is the tunable parameter whereas $\epsilon_F$ is  fixed. As all particle energies are approximately proportional to $\epsilon_B$, the energies $E_{\text{pol}/\text{mol}}$ are shifted by $\epsilon_B$ to enhance visibility. The inset shows the energies without the shift. (b) Energy spectrum expressed in units most suitable for 2D semiconductor experiments. Here  $\epsilon_B$ is fixed and  $\epsilon_F$ is varied using gate-doping. Despite the different appearance, both panels show the same data. The polaron-to-molecule transition 	is marked by the vertical, dashed gray line. The results are obtained for $h_{k=\Lambda}^2= 10^{8} \epsilon_F$ and $\Lambda^2=2.5 \times 10^5 \epsilon_F$ which ensures that the two-channel model reduces to a model of contact interactions between fermions and bosons.}\label{pmenerg} \end{figure*}

In Fig.~\ref{pmenerg} we show the polaron and the molecular energy as obtained from the two- and three-body truncations. The spectrum of the Fermi polaron problem is shown both in units convenient for cold atoms [\cref{pmenerg}(a)] as well as 2D materials [\cref{pmenerg}(b)]. The comparison of (a) and (b) demonstrates that despite the fact that both panels contain fully redundant information, they yet represent seemingly different behaviour which is, however, solely due to the different choice of units. 

In Fig.~\ref{pmenerg} the results obtained from the two-body truncation \eqref{truncation2} are shown as dashed lines.  This truncation takes into account a similar set of diagrams as a non-self-consistent $T$-matrix approach \cite{Schmidt2012} which, in turn, is equivalent to a variational Chevy approach \cite{Zoellner2011,Combescot2007}. By contrast to the aforementioned approaches our fRG is self-consistent. As expected from these approaches, we find that the two-body truncation is not sufficient to generate a polaron-to-molecule transition. 

Instead we find that the inclusion of irreducible three-body correlations is crucial, which is consistent with diagrammatic MC \cite{Kroiss2014} and higher-order variational approaches \cite{Parish2011,Parish2013}. We find that the inclusion of the three-body vertex $\Gamma_{3,k}$ lowers the molecular energy while increasing the polaron energy. As a result,
taking into account the RG flow of the irreducible atom-molecule scattering vertex $\lambda_k$  (solid lines in Fig.~\ref{pmenerg}) we find a transition at a dimensionless interaction strength $(\epsilon_B/\epsilon_F)^\ast= 18.78$ which is in excellent agreement with MC and variational results. A comparison of our result for $(\epsilon_B/\epsilon_F)^\ast$ with literature is provided in \cref{tab:critratio}.

Similar to previous field-theoretical or variational approaches \cite{Zoellner2011,Parish2011,Schmidt2012,Parish2013}, we do not include all possible two-body correlations and focus on the effect of pairing correlations. Further two-body correlations can, for instance, be generated by the re-emergence of the four-point vertex $\sim \gamma \psi^*\psi \phi^*\phi$. One may justify the exclusion of this vertex by an analogy to BEC superconductivity.  There the vertex $\gamma$ accounts for induced interactions in the particle-hole channel, leading to a contribution similar to the Gorkov corrections to BCS superconductivity \cite{gmb1961,Pethick2001,Floerchinger2008}. In the BCS case, it leads to an effective shift of the inverse dimensionless interaction strength that appears in the gap equation determining $T_c/T_F$. Based on this analogy, we expect that such terms will not  establish a new polaron-to-molecule transition, but rather  only shift the location of an already present  transition. Thus we concur with previous studies that it is  three-body correlations that are essential to establish the  formation of a phase of trions in  strongly coupled Bose-Fermi mixtures \footnote{In this argument we disregard the Coulomb repulsion between the excess charge carriers in 2D semiconductors that might further reduce the interaction range over which trions can build a stable phase.}.

We note that at low Fermi energies, we find a weak non-monotonous behaviour of the polaron energy in the dependence on $\epsilon_F/\epsilon_B$. Such a behavior is not present in  works using variational  \cite{Schmidt2012,Zoellner2011, Parish2011, Parish2013} or MC approaches \cite{Kroiss2014,Vlietinck2014}. As discussed  in \cref{app:abovebelow}, we attribute this effect to the limited resolution of the frequency- and momentum-dependence of the vertex functions in both our truncations. This effect is, however, not relevant for our study of the Bose-Fermi phase diagram which  depends only on the relative energy gaps between the polaron and molecular state and not on their respective absolute values.

\begin{table}[t]
	\centering
	\begin{tabular}{|c |  c | c|}\hline 
		Theoretical approach & $(\epsilon_B/\epsilon_F)^\ast$ \rule{0pt}{2.5ex} & $\eta_c$ \\
		\hline
		fRG (present work) & 18.78 & $-1.12$ \\
		Basic variational  \cite{Parish2011} & 9.9 & $-0.8$\\
		High-order variational  \cite{Parish2013} & 14 &$ -0.97$\\
		Diag. MC \cite{Kroiss2014} &$18.1\pm 7.2$  & $-1.1\pm0.2$\\
		Diag. MC \cite{Vlietinck2014} & $13.4\pm 4$ & $-0.95\pm 0.15$\\
		Diffusion MC \cite{Bertaina2012} & $\approx 15$ & $\approx -1$\\
		Experiment  \cite{Koschorreck2012,koehl2012} & $11.6\pm 4.6$  &$-0.88\pm0.2$\\
		\hline 
	\end{tabular} 
	\caption{Comparison of the critical ratio $(\epsilon_B/\epsilon_F)^\ast$ and the interaction parameter $\eta_c= -\log(\epsilon_B/2 \epsilon_F)/2=\log(k_F a_{2D})$ (that relates the 2D scattering length to the binding energy) obtained from  our approach (fRG, first line) with that found by previous theoretical calculations based on Monte-Carlo techniques, variational Ans\"atze and experiment. \label{tab:critratio}}
\end{table}

\subsection{Vertex functions}

\begin{figure}[t]
	\begin{center}
		\includegraphics[width=\linewidth]{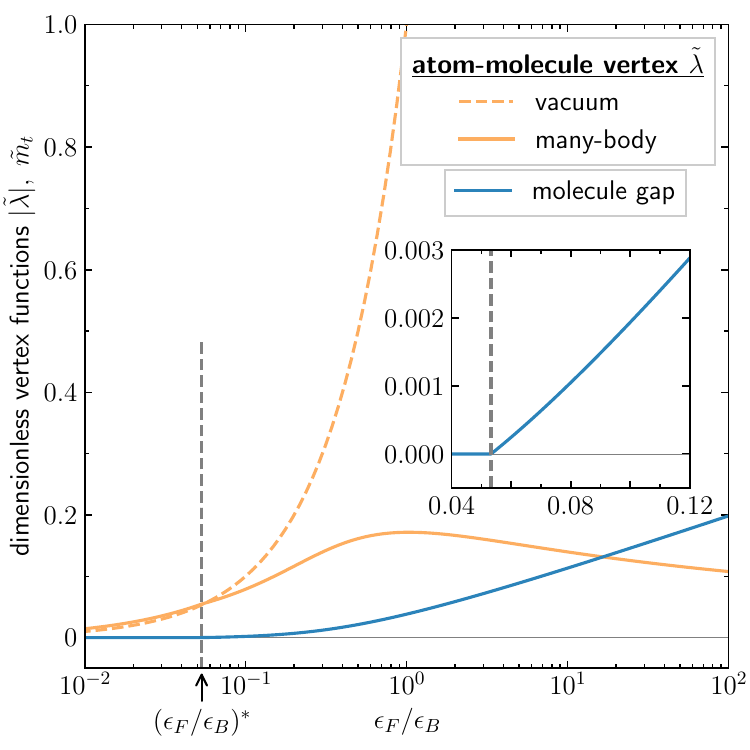}
	\end{center}
	\caption{Renormalized dimensionless three-body vertex  $|\tilde \lambda|=\left|\lambda_{k=0}\epsilon_F /h_{k=0}^2\right|$ (orange, solid)  and 
	renormalized dimensionless molecular gap  $\tilde{m}_{t}=m_{t,k=0}/h_{k=0}^2 $ (blue, solid) obtained within the three-body truncation $\Gamma_{3,k}$ as a function of $\epsilon_F/\epsilon_B$. The dashed orange line shows the value of the  atom-molecule scattering vertex in the three-body (vacuum) limit where $\tilde\lambda\to \tilde \lambda^{(3B)}=-\epsilon_F/ \epsilon_B$. 
	 Deep in the strong-binding or, equivalently, low fermion doping regime $\epsilon_F/\epsilon_B\ll1$, $\tilde \lambda$ approaches the three-body result.  Starting at around the scale $\epsilon_F\approx \epsilon_B$, medium corrections to the atom-molecule scattering lead to a pronounced suppression effect.
		The inset shows that close  to the polaron-to-molecule transition the molecular gap $\tilde{m}_{t}$ 
vanishes linearly as $(\epsilon_F/\epsilon_B)-(\epsilon_F/\epsilon_B)^\ast $. In the two-body truncation $\lambda_k\equiv 0$, and $\tilde{m}_{t}$ remains positive for all $\epsilon_F/\epsilon_B$, since no polaron-to-molecule transition exists at this level of approximation. The results are obtained for $\Lambda^2=2.5 \times 10^5 \epsilon_F$ and $h_{k=\Lambda}^2= 10^{8} \epsilon_F$.}\label{vertexvalues} \end{figure}

The FRG approach allows one to analyze the two- and three-body vertices that determine the emergent effective interactions and correlations  in the system.
In Fig.~\ref{vertexvalues}, the dimensionless, renormalized  atom-molecule scattering vertex  $\tilde \lambda \equiv\lambda_{k=0} \epsilon_F/h^2_{k=0}$ and molecular gap $\tilde m_t \equiv m_{t,k=0}/h_{k=0}^2$ are shown as function of $\epsilon_F / \epsilon_B$. We have scaled both vertices  by  powers of $h$ that reflect the scaling of the vertices with the molecular wave function renormalization  $A_{t,k=0}^{-1}$ yielding results independent of  $h$ in the contact-interaction limit at $h\to\infty$.

\textbf{Atom-molecule  scattering.---}
The vertex $\lambda_k$ describes the scattering between the composite fermionic molecules and the excess fermions in the system. During the RG flow, $\lambda_k$ evolves from $\lambda_{k=\Lambda}=0$ in the UV to a  negative value in the infrared at $k=0$. Thus $\tilde \lambda$ yields  an attractive contribution, shown in \cref{fig:treelevelthreebody}(c), to the overall atom-molecule scattering amplitude that has an additional, significant contribution from the tree-level $\phi$-exchange diagram depicted in \cref{fig:treelevelthreebody}(d). 

 Fig.~\ref{vertexvalues}  shows the absolute value of the scattering vertex in the three-body limit (dashed orange line) where it takes the value $\tilde \lambda=\tilde \lambda^{(3B)}=-\epsilon_F/\epsilon_B$,  
 for details see \cref{threebodyvacuum}. Thus the vertex scales proportional to the square of the size of the molecular bound state $a_B\propto \sqrt{1/\epsilon_B}$.
The solid orange line shows the result for $\tilde \lambda$ in the polaron problem. At small fermion density the molecule is the ground state.  In this \emph{`molecular regime'} the density of fermions is so low that the average inter-fermion spacing greatly exceeds the molecular size $a_B$. Thus the atom-molecular scattering vertex is essentially unaffected by the presence of the fermionic medium, and $\tilde \lambda$ follows the three-body result $\tilde \lambda^{(3B)}$. 

As $\epsilon_F/\epsilon_B$ is increased we observe a suppression of the atom-molecule scattering vertex. We attribute this effect to two contributing factors. First,  the molecule becomes an excited state beyond the critical interaction $(\epsilon_F/\epsilon_B)^*$. In this case the molecule is gapped and within our FRG approach which  projects vertex functions on vanishing external vertex frequencies and momenta (see \cref{sect:AppFloweqs}), $\tilde{\lambda}$ is thus suppressed by the molecular energy gap. More importantly, however, as the Fermi energy becomes larger than $\epsilon_B$, $\epsilon_F/\epsilon_B>1$, the size of the bound state starts to exceed the typical inter-fermion distance. As a consequence, in-medium effects come into play  leading to significant modifications of $\tilde \lambda$. Indeed, these corrections become so strong  that  $\tilde{\lambda}$  starts to decrease at even larger values of $\epsilon_F/\epsilon_B$.

\textbf{Molecular gap.---}  The dimensionless molecular gap $\tilde m_t=m_{t,k=0}/h_{k=0}^2 $ is shown as a blue line in  Fig.~\ref{vertexvalues}. For interaction strengths $\epsilon_F/\epsilon_B<(\epsilon_F/\epsilon_B)^*$ where the molecule is the ground state, the molecule is gapless, $\tilde m_t = 0$. Beyond the transition the molecule becomes an excited state and we find that its gap vanishes linearly as $m_t\sim(\epsilon_F/\epsilon_B)-(\epsilon_F/\epsilon_B)^\ast $ towards the transition. 

The corresponding crossing of the molecular and the polaron state can also be interpreted as leading to an effective Feshbach resonance in the polaron-fermion scattering where the tree-level diagram shown in \cref{fig:treelevelthreebody}(b), evaluated  on-mass-shell, diverges. The associated polaron-fermion scattering length changes sign at the transition, with a positive value signaling the existence of a fermionic bound state. 

In turn, within a single-channel theory that is  formulated purely in terms of the `atomic fields' $\psi$ and $\phi$, the divergence of the effective polaron-fermion scattering vertex $\sim h^2/P_t$  signals the instability towards a phase of fermionic bound states. In this language, entering this phase at finite boson density would necessarily require the introduction of the emergent fermionic composite states. Finally we note that in Fig.~\ref{vertexvalues} we show only results from the three-body truncation $\Gamma_{3,k}$ since in the two-body truncation $\Gamma_{2,k}$, the vertex $\lambda_k=0$  vanishes by definition throughout the RG flow. Moreover, since no polaron-to-molecule transition is present in this simpler truncation, $\tilde m_t$ always remains finite.

\section{Bose-Fermi mixture at finite boson density}
\label{sect:finitedens}

\begin{figure*}[t]
  \begin{center}
		\includegraphics[width=\linewidth]{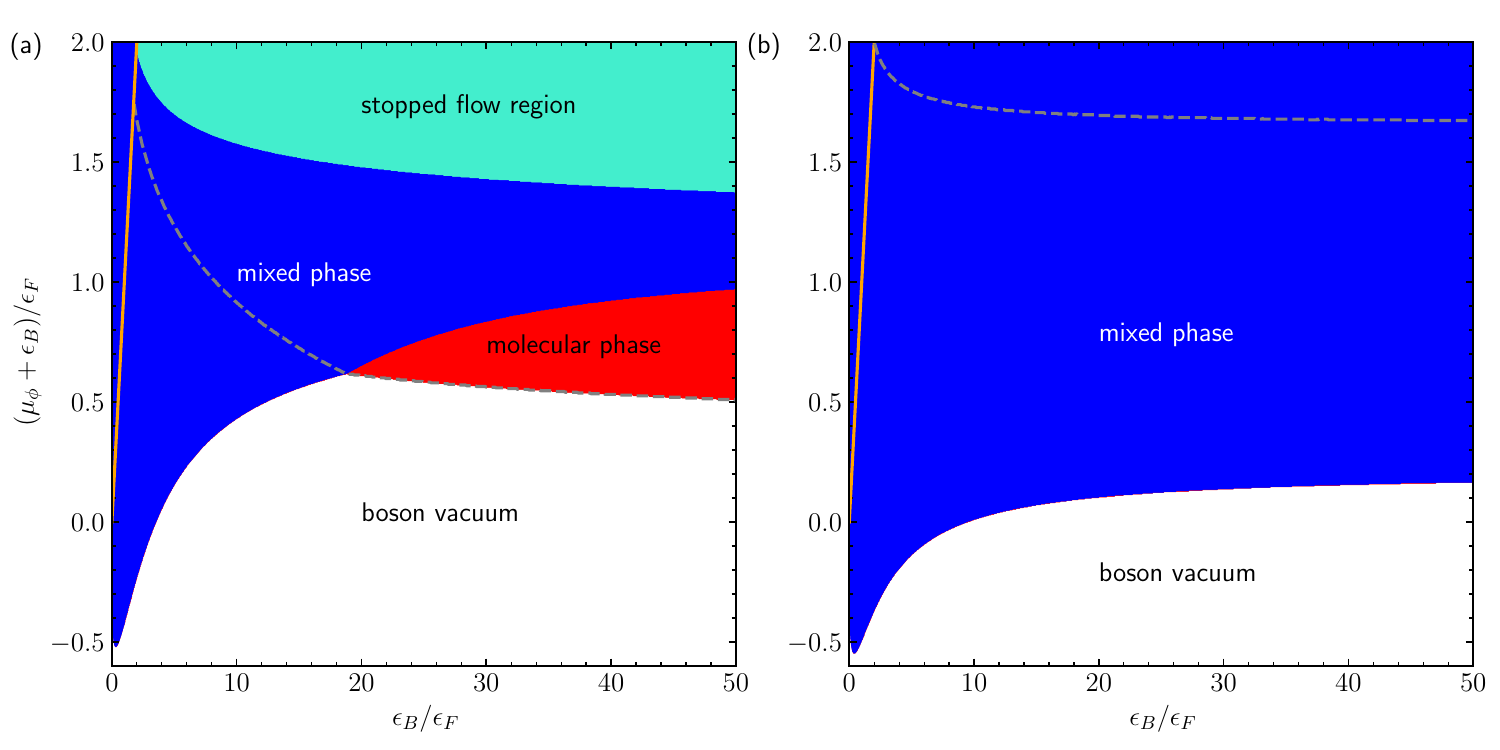}
  \end{center}
	\caption{Phase diagram of the Bose-Fermi mixture for different ratios of $\mu_{\phi}/\epsilon_F$ and $\epsilon_B/\epsilon_F$ using $\Gamma_{3,k}$ (a) and $\Gamma_{2,k}$ (b). The white regions indicate the vacuum phase, while the red and blue regions denote the \emph{molecular} and the \emph{mixed} phase. The mint-colored region above the mixed phase denotes the \emph{stopped flow region} in which the flow was stopped at $k>0$ because the molecular Fermi surface became larger than the majority's $(-2 m_{t,k}/A_{t,k}>\epsilon_F)$, indicating the breakdown of our approximation. The gray dashed line denotes the boundary above which the molecules form a Fermi surface at the end of the flow, i.e.  $m_{t,k=k_\text{end}}/A_{t,k=k_\text{end}}<0$. The orange line indicates the path along which $\mu_\phi=0$. In both truncations  the boundary between the vacuum  and finite density phases approaches the origin for $\epsilon_B/\epsilon_F\to 0 $.}\label{chemicalpot} \end{figure*}

We now turn to the mixture regime, where a finite density of bosons interacts with a bath of majority fermions. As discussed in \cref{sec:truncation}, within our truncations we can identify two phases:  a molecular phase, where all bosons are bound into  molecules, and a mixed phase where molecules are hybridized with majority fermions and coexist with a condensate of polarons. 

While we can describe the molecular phase directly, we can not fully access the regime in which a condensate of polarons exists since this would require to explicitly include the condensate and thus an effective potential for the bosonic field. However, we can still  determine the critical system parameters at which the system becomes unstable towards condensation. Indeed, the associated phase boundary is determined by the vanishing of the scale-dependent boson gap $m_{\phi,k}/A_{\phi,k}$ at the end of the RG flow.

For large values of the boson chemical potential $\mu_\phi$, the underlying assumption $n_B \ll n_F$ is no longer valid. When this condition breaks down, we thus terminate the fRG flow. While this does not define a phase, we dub this part of the phase diagram the \emph{`stopped flow region'}, further discussed below.

\subsection{Phase diagram as a function of chemical potential}

In Fig.~\ref{chemicalpot} we present the phase diagram of the Bose-Fermi mixture  for both $\Gamma_{2,k}$ and $\Gamma_{3,k}$ [\cref{truncation2,truncation3}]  at a fixed  Fermi energy $\epsilon_F$, as function of $\epsilon_B$ and $\mu_{\phi}$. 
In the three-body truncation $\Gamma_{3,k}$ [Fig.~\ref{chemicalpot}(a)] a molecular phase forms at finite boson density in the interaction regime where the molecule is the ground state of the quantum impurity limit discussed in Section~\ref{sect:polaron}.

In fact, the ground-state energy of the quantum impurity limit determines the chemical potential $\mu^{\phantom{c}}_{\phi}=\mu_{\phi}^c(\epsilon_B,\epsilon_F)$ that separates the vacuum of bosons  from the mixed  phase  or the phase of a finite density of molecules. Along this phase boundary the system undergoes a transition from a polaronic to a molecular ground state.

In the interaction regime $\epsilon_B/\epsilon_F>(\epsilon_B/\epsilon_F)^*$,  increasing the boson chemical potential starting from  values $\mu_\phi<\mu_{\phi}^c$ leads  to a boson-vacuum-to-molecule transition as $\mu_\phi$ crosses the critical chemical potential. Directly on the critical line one enters the quantum impurity regime and a single molecule forms \footnote{Strictly speaking along the critical line any \textit{finite} particle number can be realized as long as the boson density $n_B$ vanishes in the thermodynamic limit. In a field theory approach the exact particle number considered is then determined by the highest-order vertex function taken into account.}. Increasing $\mu_\phi$ beyond $\mu_\phi^c$ one enters the molecular phase where a finite density of  bosons, all bound into molecules, exists. In this phase $m_{t,k=0}/A_{t,k=0}<0$, and the molecules acquire a Fermi surface. Tuning $\mu_\phi$ further to larger values one  reaches the phase boundary to the mixed phase. Here, a finite density of molecules coexists with gapless boson particles. 

For $\epsilon_B/\epsilon_F<(\epsilon_B/\epsilon_F)^*$ there is no molecular phase and  one transitions directly from the boson vacuum to the mixed phase. As the flow is terminated at a finite RG scale $k_\text{end}$ once the boson becomes gapless $m_{\phi,k=k_{\text{end}}}/A_{\phi,k=k_{\text{end}}}=0$, at the boundaries to the molecular phase and to the boson vacuum phase the boson turns gapless at the end of the flow at $k_{\text{end}}=0$. Moving further into the phase from these boundaries the value of $k_{\text{end}}$ at which the flow is terminated increases. 

When the flow is stopped in the mixed phase, the molecules might have already formed a molecular Fermi level during the course of the RG flow. This is indicated by the  dashed gray line in \cref{chemicalpot}. Above this line the molecule has developed a Fermi surface when the flow ends or is terminated at $k=k_{\text{end}}$. Below the line the molecule has remained gapped. As expected, for $\epsilon_B/\epsilon_F>(\epsilon_B/\epsilon_F)^*$ this line parametrizes the boson-vacuum-to-molecule transition. For $\epsilon_B/\epsilon_F<(\epsilon_B/\epsilon_F)^*$ on the other hand, it bisects the mixed phase. These regions then correspond to phases of a single Fermi sea (boson vacuum), two Fermi seas (molecular phase), two Fermi seas with a bosonic condensate (mixed phase above the gray dashed line) and a bosonic condensate with only a single Fermi sea (mixed phase below the gray dashed line) as discussed in Refs.~\cite{Yabu2003,Powell2005}.

Increasing the bosonic chemical potential $\mu_\phi$ further within the mixed phase, the bosonic density increases until eventually the molecular Fermi wave vector becomes larger than the fermionic Fermi wave vector ($-2 m_{t,k}/A_{t,k}>\epsilon_F$). Within this regime, the bosonic density has become comparable to the fermionic density. 
This means that it is no longer justified to neglect the renormalization of the fermionic Green's function and to disregard higher-order correlations along with sub-dominant interaction channels. As we expect that in this case our truncation no longer renders an appropriate description of the system, we terminate the flow  at finite scale $k_{\text{end}}$ once $-2 m_{t,k}/A_{t,k}>\epsilon_F$. 
When this happens \textit{during} the RG flow, a molecular Fermi sea has already formed while the bosons are still gapped $m_{\phi,k}/A_{\phi,k}>0$. 
This \emph{`stopped flow region'} (mint in \cref{chemicalpot}) occurs after the bosonic chemical potential has been tuned well into the mixed phase. We therefore expect that in the stopped flow region, close to the boundary to the mixed phase, the system would still be in a mixed phase, if one were to continue the flow.

Within the two-body truncation [see Fig.~\ref{chemicalpot}(b)] it is unsurprising to see that no molecular phase forms at finite boson density, since already in the single-boson regime this \emph{Ansatz} does not form a molecule in the ground state. Rather, one transitions from the boson vacuum phase directly to the mixed phase as the molecule only becomes gapless at $k_{\text{end}}$ well within the mixed phase (gray dashed line). Within this truncation the stopped flow regime is not realized for the considered range of $\mu_\phi$ and it sets on only at around $(\mu_\phi+ \epsilon_B) \approx 2.4 \epsilon_F$.

\subsection{Phase diagram as a function of density}
\label{sect:FlowingDens}
In the previous subsection results were given as a function of chemical potential. Experimentally it is, however, often simpler to determine the  density of particles instead of their chemical potential. Thus, to make direct connection to experiments, it is useful to also consider the phase diagram  as a function of particle densities. Since in the effective action formalism employed in this work, the chemical potentials are the parameters of the theory, the canonically conjugate densities have to be computed explicitly.

In principle, the  fermion and boson densities can be determined directly from the two-point Green's functions. Within the derivative expansion and two-channel model an alternative approach is, however, more convenient. Here one makes use of the fact that the densities are connected to the derivative of the effective potential $U$ evaluated at the equilibrium field configuration $\sigma_{\text{eq}}$ by the standard relation
\begin{equation}
n_{F/B}=-\dfrac{\partial U(\sigma_{\text{eq}})}{\partial \mu_{\psi/\phi}} \ .
\label{eq:defNFB}
\end{equation}
Here $n_{B}$ and $n_F$, respectively, denote the total density of bosons and fermions in the system, including those bound into molecules. The effective potential $U(\sigma_{\text{eq}})$, in turn, is obtained from the derivative-free part of the infrared effective action evaluated at the field expectation values $U(\sigma_{\text{eq}})= \Gamma_{k=0}[\sigma_{\text{eq}}]/(V/T)$ where for the considered phases $\sigma_{\text{eq}}=(\psi_{\text{eq}},\phi_{\text{eq}},t_{\text{eq}})=0$.

In the absence of approximations, determining the densities from the effective potential or from the Green's functions are equivalent methods, as follows from the \emph{Luttinger} theorem \cite{Powell2005,Abrikosov1975}. Within our fRG scheme we, however, expect it to be computed more accurately using the flow of $U$ than using the flow of $G_{\sigma}$ as this approach relies on lower-order vertices.

In the fRG, the effective action is promoted to a flowing effective action that depends on the RG scale $k$. Accordingly, it is convenient to define corresponding scale-dependent densities $n_{F/B,k}$ and to determine the densities of the systems from their value at the end of the RG flow. The resulting density values are then associated with the corresponding phases. Since there is no polaron-to-molecule transition for $\Gamma_{2,k}$, in the following we discuss only results obtained in the three-body truncation.

The flow equation of the effective potential is  obtained by evaluating the Wetterich equation \eqref{wetterich} at vanishing fields,
\begin{align}
\partial_k U_k(\sigma_{\text{eq}})= \sum_{\sigma=\psi,\phi,t}\xi_\sigma \int_{P} G^{c}_{\sigma,k}(\pv, \omega) \partial_k R_{\sigma,k}(\pv,\omega) \label{effpotflow1}
\end{align}
where $\xi_\phi=1$ for bosons and $\xi_\sigma=-1$ for fermions ($\psi$ and $t$). Due to the pole structure of the integrand, \cref{effpotflow1} can be simplified further (for details see \cref{sect:effPot}) to  
\begin{align}
\partial_k U_k(\sigma_{\text{eq}})=& -\int_{P} \left[1-\Theta_{\psi,k}(\pv) \right] G_{\psi,k}(\pv,\omega) \partial_k G^{-1}_{\psi,k}(\pv,\omega) \nonumber\\
&+\int_{P} \left[1-\Theta_{\phi,k}(\pv) \right] G_{\phi,k}(\pv,\omega) \partial_k G^{-1}_{\phi,k}(\pv,\omega) \nonumber\\
&-\int_{P} \left[1-\Theta_{t,k}(\pv) \right] G_{t,k}(\pv,\omega) \partial_k G^{-1}_{t,k}(\pv,\omega) \ . \label{partialU}
\end{align}
Here, the step functions $\Theta_{\sigma,k}(\pv)$ originate  from the sharp regulators in the flow equations [see \cref{regulatorPsi,regulatorPhi,regulatorT}]. For the bosonic and molecular field they are defined as $\Theta_{t,k}(\pv)=\Theta_{\phi,k}(\pv)=\Theta(\pv^2-k^2)$, while for the fermionic field $\Theta_{\psi,k}(\pv)$ is defined in the following in \cref{fermionstep}.   

As the scheme described in \cref{sect:Model} does not feature
a renormalization of the majority propagator it is evident from \cref{partialU} that, within that approximation, the fermions do not contribute to the flow of the effective potential $U_k$. Consequently, from the integration of \cref{partialU} the density of fermions would not be calculated accurately since the depletion of majority carriers, resulting from fermions being bound into molecules, is not taken into account.

In order to take this effect into account, we derive ---\emph{separate} from the flow of the Green's functions of the bosons, molecules and the interaction vertices--- a flow equation for the propagator of the majority species, that does not feed back into any flow other than that of the effective potential. Since the majority fermions have a finite density already at the start of the RG flow, we regulate the fermions around their flowing Fermi level $\epsilon_{F,k}= \epsilon_F- m_{\psi,k}/ A_{\psi,k}$ \cite{Floerchinger2010}. Accordingly, the step function in the first line in \cref{partialU} is given by
\begin{align}
\Theta_{\psi,k}(\pv)= \Theta (|\pv^2 - \epsilon_F+ m_{\psi,k}/A_{\psi,k}|-k^2) \ . \label{fermionstep}
\end{align} 
To derive the  flow equations of $A_{\psi,k}$ and $m_{\psi,k}$, we evaluate the RG flow of the associated vertex function at external frequency and momentum $(\pv^2,\omega)=(\epsilon_F,0)$, i.e., we perform the gradient expansion around the bare Fermi surface of the majority species. This flow is  then used to determine the effective  potential $U$, and, in turn, the boson and fermion densities through \cref{eq:defNFB}. In order to reproduce the majority carrier density $\epsilon_F/4\pi$ in the UV with regard to \cref{eq:defNFB}, the initial condition for the density flow is given by the mean-field result $U_{k=\Lambda}(\sigma_{\text{eq}})= -\epsilon_F^2/8 \pi$.

\begin{figure}[t]
\begin{center} 
\includegraphics[width=\linewidth]{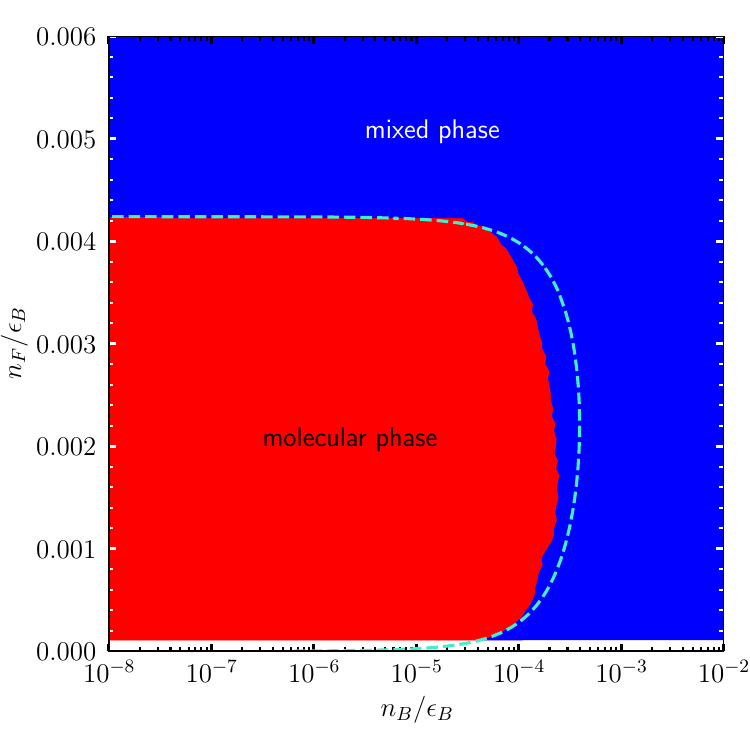} 
\end{center}
\caption{Phase diagram of the Bose-Fermi mixture for different boson and fermion densities  $n_F$ and $n_B$ at a fixed interaction strength set by $\epsilon_B$. The red region indicates the molecular phase while the blue region indicates the mixed phase. The mint-colored dashed line denotes the mean-field phase boundary extrapolated from the Fermi polaron problem. 
}\label{densdens}
\end{figure}

In \cref{densdens} we show the resulting phase diagram of the system as a function of the boson and fermion density. It can be regarded as the counterpart of \cref{chemicalpot}(a), expressed in different variables. Specifically, to obtain \cref{densdens}, for the combinations of boson chemical potential $\mu_\phi$ and interaction strength $\epsilon_B/\epsilon_F$ that lie in the molecular phase we computed the corresponding values of $n_B/\epsilon_B$ and $n_F/\epsilon_B$. For combinations that lie inside the mixed phase or the stopped flow region we can not compute the boson and fermion density as the flow is terminated at finite $k_\text{end}$. In \cref{densdens} we thus identify density combinations outside the molecular phase as being in the mixed phase \footnote{Fig.~\ref{densdens}  only shows density ratios of the mixed phase in vicinity of the molecular phase. Since in \cref{chemicalpot}(a) the stopped flow region does not border the molecular phase directly, we thus do not expect the stopped flow region to appear close to the molecular phase in \cref{densdens}.}.

In Fig.~\ref{densdens}, the single-boson limit discussed in \cref{sect:polaron} corresponds to the $y$-axis at $n_B/\epsilon_B=0$, and the polaron-to-molecule phase transition occurs at $n_F/\epsilon_B =(\epsilon_F/\epsilon_B)^* /4\pi=0.00424$. As the boson density is increased, the mixed phase becomes  favorable, i.e., the \emph{maximal} density of fermions for which all bosons are bound into molecules decreases. We find that there is also a \emph{minimal} fermion density required to enter the molecular phase. Below that critical value one again enters the mixed regime.

\subsection{Mean-field model}

Remarkably, a simple mean-field-inspired argument can provide an approximate phase diagram of the model: in the  single-boson limit, the polaron is a gapped excitation in the molecular regime. It has a gap $\Delta E = E_\text{pol}-E_\text{pol}$ which is a function of $\epsilon_F/\epsilon_B$, or equivalently $n_F/\epsilon_B$ (equal to $\epsilon_F/4\pi\epsilon_B$ along the y-axis in \cref{densdens}). This gap was determined numerically in \cref{sect:polaron} where we found,
\begin{align}
\Delta E ((\epsilon_F/\epsilon_B)^*)={}&0,&
\Delta E ((\epsilon_F/\epsilon_B) \to 0)\approx {}&0.41 \epsilon_F ,
\end{align}
reflecting that the energy gap vanishes at the polaron-to-molecule transition and attains a value proportional to $\epsilon_F$ in the strong-binding, low-density limit.

In our mean-field model of the molecular phase, the interactions are taken into account by considering the  effective Hamiltonian 
\begin{align}\label{MFModel}
H^{MF}= \sum_\veck & \Big[\varepsilon_\veck \psi^{\dagger}_{\veck}\psi^{\phantom{\dagger}}_{\veck} + \frac{\varepsilon_\veck}{2}  t^{\dagger}_{\veck}t^{\phantom{\dagger}}_{\veck}\\ \nonumber
&+ (\varepsilon_\veck + \Delta E (\epsilon_F/\epsilon_B)) \phi^{\dagger}_{\veck}\phi^{\phantom{\dagger}}_{\veck}\Big]
\end{align}
where $\varepsilon_\veck= \veck^2 / 2m $.
Even though $H^{MF}$ is quadratic in the fields, this effective model goes beyond naive mean-field as $\Delta E$ incorporates the non-trivial solution of the polaron problem obtained through our fRG scheme in \cref{sect:polaron}. The polaronic, mixed phase appears when it is energetically unfavorable to bind into molecules, i.e. when the Fermi energy $\epsilon_{F,t}$ of the molecules is larger than the gap $\Delta E$. When this condition is reached the polarons start to form a condensate, as  described previously. 

For a molecular Fermi energy below the gap $\Delta E$, the ground state of the mean-field model \eqref{MFModel}  is given by separate Fermi seas of densities $n_\psi=\epsilon_F/4\pi$ and $n_t= \epsilon_{F,t}/2 \pi$ for the fermionic and molecular sectors, respectively. Hence, in the molecular phase, the \emph{total} bosonic and  fermionic densities are given by $n_B=n_\phi+n_t=n_t$ and $n_F=n_\psi+ n_t= n_t+\epsilon_F/4\pi$. The mean-field transition line below which the molecular state is favored is thus parametrized by
\begin{align}
\dfrac{n_B}{\epsilon_B} &= \dfrac{\Delta E(\epsilon_F/\epsilon_B)}{2 \pi \epsilon_B} \ ,\\ \frac{n_F}{\epsilon_B}&= \frac{\epsilon_F}{4\pi \epsilon_B}+  \frac{\Delta E(\epsilon_F/\epsilon_B)}{2 \pi \epsilon_B} \ .
\end{align}

This mean-field phase boundary is shown as a dashed line in \cref{densdens}. While the mean-field picture is over-simplified and does not correctly capture the quantitative renormalization effects beyond the vacuum-to-molecule transition, it correctly captures the qualitative nature of the structure of the phase diagram. The phase boundary, by construction, reaches the y-axis at the polaron-to-molecule transition and approaches the origin at an angle of about $n_F/n_B\approx  2.22 $  which directly follows from the behaviour of the polaron gap $\Delta E((\epsilon_F/\epsilon_B) \to 0)\approx 0.41 \epsilon_F$.

\section{Quasiparticle properties of polarons and molecules in the quantum impurity limit}
\label{sect:Kamikado}

The calculations presented in \cref{sect:polaron,sect:finitedens} only yield information about ground state properties of the system. In order to extract spectral information such as dispersion relations, particle lifetimes, effective masses or higher-lying excited states, however, the spectral functions need to be computed. 

The spectral functions are obtained from the Green's functions $G_{\psi,\phi,t}$ by analytic continuation of the Matsubara frequencies $i \omega \to \Omega + i 0^+$ which yields the retarded Green's functions $G^R_{\psi,\phi,t}(\pv,\Omega)$. From this, the momentum- and frequency-resolved spectral functions  are obtained as
\begin{align}
\Ac_{\psi,\phi,t}(\pv,\Omega)= \operatorname{Im} \dfrac{1}{\pi} G_{\psi,\phi,t}^R{}(\pv,\Omega)\ . \label{eq:defSpecFun}
\end{align} 

\begin{figure*}[t]
	\begin{center}
		\includegraphics[width=\linewidth]{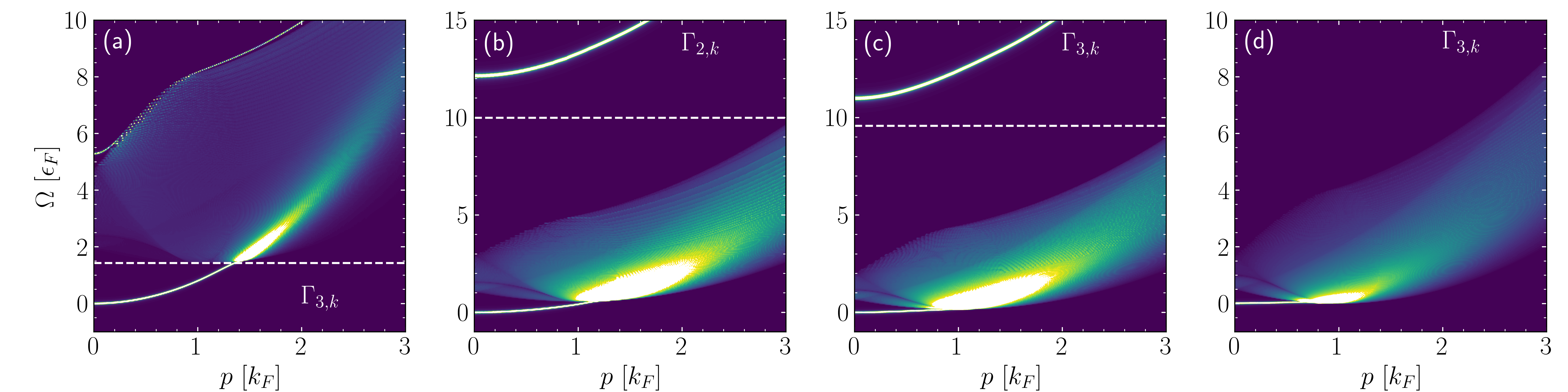}
	\end{center}
	\caption{Polaron spectral function $\Ac_{\phi}$ for different dimensionless interaction strengths $(\epsilon_B/\epsilon_F)$: (a) 1, (b) 10, (c) 10, (d) 20. In (a), (c) and (d) $\Gamma_{3,k}$ is used while in (b) $\Gamma_{2,k}$ is used to highlight the effect of the renormalization of the three-body sector. Dashed horizontal lines denote the Bose-Fermion scattering threshold at $\Omega=-\mu_\phi$. In (a) the range of the color spectrum is $[0,0.5]$, while in (b), (c), (d) it is $[0,0.05]$. }\label{KamikadoPolaron} \end{figure*}

Two difficulties arise when determining the spectral function within the fRG. First, an analytic continuation has to be performed, either at the level of the flow equations~\cite{Floerchinger2012,Pawlowski2015,Kamikado2013} or the final output of the RG flow in the infrared \cite{Schmidt2011}. Second, in order to capture non-trivial spectral functions one needs the full momentum- and frequency-dependence of the propagator, which the gradient expansion  employed in \cref{sect:polaron,sect:finitedens} does not provide. A solution to the latter difficulty can be found, e.g.,  by the direct implementation of  fully frequency- and momentum-resolved Green's functions \cite{Schmidt2011} or in the BMW scheme~\cite{Blaizot2006,Benitez2012}, which also yields a full momentum- and imaginary frequency-dependence of the propagators. Both these approaches, however,  do not resolve the analytic continuation issue. For this reason, we  implement here a method developed in nuclear physics \cite{Kamikado2013,Kamikado2014,Tripolt2014,Tripolt2014a} which was recently applied to the polaron problem in three dimensions  \cite{Kamikado2017}. In the following we shall refer to this method as the frequency- and momentum-resolved scheme (FMR).

In FMR, the flow equations [\cref{wetterich} and \cref{Gphiflow,Gpsiflow,Gtflow,hflow,lambdaflow}] are analytically continued to real frequencies. In order to achieve that, rather than projecting the flow equation onto the gradient expansion parameters, we retain the full momentum- and frequency-dependence of the single-particle Green's functions on the lhs. of the flow equations, while we keep the gradient expansion for the two-body [\cref{truncation2}] and three-body truncation [\cref{truncation3}] on the rhs. of the equations. This enables us to perform the loop integration over imaginary frequencies analytically. In turn, this allows us to perform  the analytic continuation to real frequency to obtain direct access to the retarded Green's functions. From that we evaluate the single-particle spectral function using \cref{eq:defSpecFun}; for further details we refer to \cref{sect:AppImag}. We remark that, when applying a non-self-consistent implementation of FMR ---in which only bare quantities appear on the rhs. of the flow equations--- to the spectral function of the molecule,  the differential equation system yields the same results as a corresponding $T$-matrix resummation \cite{Schmidt2012} (see \cref{App:Equivalence}).

\textbf{Polaron spectral function.---} The polaron spectral function obtained using FMR is shown for different interaction strengths in \cref{KamikadoPolaron}. Subfigures (a), (c) and (d) are obtained in the three-body truncation $\Gamma_{3,k}$.  Subfigure (b) shows the result from the two-body truncation $\Gamma_{2,k}$  in order to highlight the effect of the inclusion of irreducible three-body correlations. 

The  polaron spectral functions show the same qualitative behavior as the corresponding spectra in 3D~\cite{Schmidt2011, Kamikado2017}.  Two quasiparticle peaks ---the attractive and the repulsive polaron---  can be discerned, and a molecule-hole continuum in between these dominant excitations is visible. The attractive polaron is the ground state in \cref{KamikadoPolaron} (a), (b), (c), and thus is a gapless  excitation. In contrast, in \cref{KamikadoPolaron}(d),  the ground-state is a molecule, and thus   a small gap at $p=0$ can be seen. Generally, at finite but small momenta the attractive polaron  is a well-defined quasiparticle with an interaction-dependent effective mass which, along with the effective masses of the repulsive polaron and the molecule, is shown in \cref{effectivemass}. For larger momenta, the attractive polaron peak eventually merges with the molecule-hole continuum, such that it is no longer a well-defined quasiparticle.

\begin{table}
		\begin{center}
		\begin{tabular}{|c | c c c | c c c | c c c|}\hline
	 & \multicolumn{3}{c|}{att. Pol.} & \multicolumn{3}{c|}{rep. Pol.}& \multicolumn{3}{c|}{Mol.}   \\
			$(\epsilon_B/\epsilon_F)$ & $\Gamma_{3,k}$&& $\Gamma_{2,k}$&$\Gamma_{3,k}$& & $\Gamma_{2,k}$ &$\Gamma_{3,k}$& & $\Gamma_{2,k}$  \\
			\hline
			1 & 0.62 & & 0.6 & 0.05 & &.056& $-.11$& &$-.09$\\ 
			10 & 2.29 & & 1.2 & 0.3& & 0.42&$\geq 6$& &$-.69$ \\ 
			20 & $\approx5$& &1.42 & 0.32& & 0.46 & 1.84& &$-1.04$\\ 
			\hline
		\end{tabular} 
	\end{center}
	\caption{Effective masses of the attractive and repulsive polaron as well as the molecule obtained from quadratic fits to the dispersion relation at $p=0$, both in the three-body and the two-body truncation.} \label{effectivemass}
\end{table}

The repulsive polaron appears at energies above the scattering threshold (indicated by the dashed horizontal lines in \cref{KamikadoPolaron}) as a narrow peak, indicating a long quasiparticle life-time for the interaction strengths shown. Consistent with Ref.~\cite{Schmidt2012} we find that as $\epsilon_B/\epsilon_F$  decreases, the repulsive polaron  gradually disappears.  Moreover, while at small interaction strength (Fig.~\ref{KamikadoPolaron}(a)) the repulsive polaron eventually merges with the molecule-hole continuum at finite momentum, at larger interaction strength the repulsive polaron peak remains distinct from the molecule-hole continuum at any momentum and thus keeps a long life-time  at high momenta. 

As evident from the comparison of \cref{KamikadoPolaron}(b) and (c), the inclusion of the irreducible three-body correlations moves the molecule-hole continuum to lower energies. This has the effect that the dispersion relation of the attractive polaron becomes flatter, increasing the polaron effective mass compared to the two-body truncation (see \cref{effectivemass}). Furthermore, its quasiparticle peak joins the continuum at lower momenta. In \cref{repulsivepolaronenergy} the energy of the repulsive polaron is shown relative to the ground-state energy. For the repulsive polaron the inclusion of three-body correlations has the effect of slightly altering its effective mass and of lowering  its energy relative to the scattering threshold.  For a fermionic impurity this  indicates a reduced tendency towards itinerant Stoner ferromagnetism \cite{Massignan2011}. 

\begin{table}[b]
	\begin{center}
		\begin{tabular}{|c | c|  c |}\hline
			$(\epsilon_B/\epsilon_F)$\rule{0pt}{2.5ex}     & rep Pol. $\Gamma_{2,k}$ & rep. Pol. $\Gamma_{3,k}$ \\
			\hline
			1 & $E_{\text{pol}}+\ 5.67 \epsilon_F$&$E_{\text{pol}}+\ 5.29 \epsilon_F$ \\ 
			10 &$E_{\text{pol}}+12.15 \epsilon_F$ &$E_{\text{pol}}+10.99 \epsilon_F$\\ 
			20 &$E_{\text{pol}}+21.92 \epsilon_F$ &$E_{\text{mol}}+20.52 \epsilon_F$\\ 
			\hline
		\end{tabular} 
	\end{center}
	\caption{Energy of the repulsive polaron at different interaction strengths obtained in the two-body and three-body truncation. The energies are given with respect to the respective ground-state energies, $E_{\text{pol}}$ and $E_{\text{mol}}$.} \label{repulsivepolaronenergy}
\end{table}

\begin{figure}[ht]
	\begin{center}
        \includegraphics[width=\linewidth]{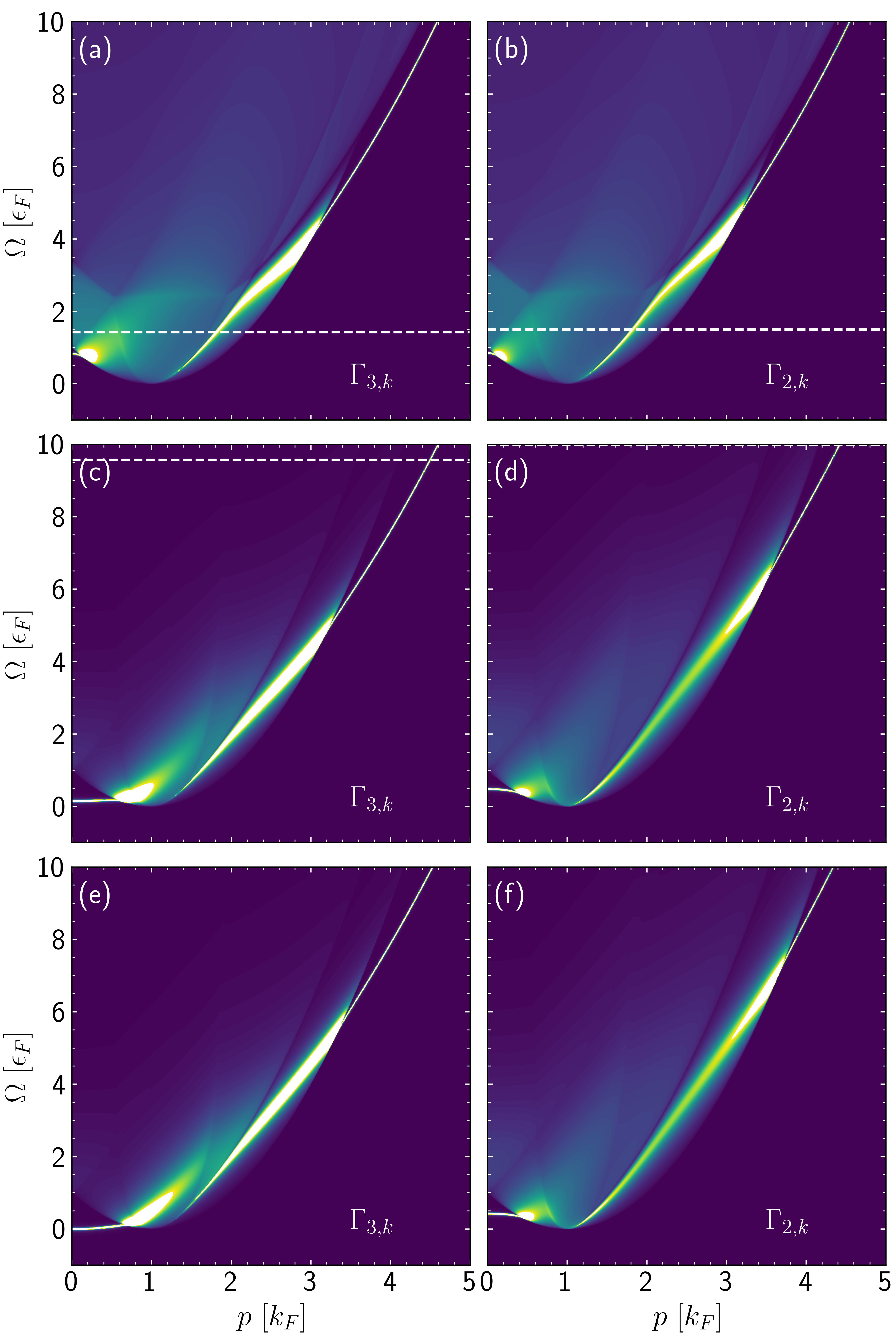} 
	\end{center}
	\caption{Molecular spectral function $\Ac_{t}$ for different dimensionless interaction strengths $\epsilon_B/\epsilon_F$: (a), (b), 1, (c), (d) 10, (e), (f) 20. In (a), (c) and (e) $\Gamma_{3,k}$ is used while in (b), (d) and (f) $\Gamma_{2,k}$ is used to highlight the effect of the renormalization of the three-body sector. Dashed horizontal lines denote the Bose-Fermion scattering threshold given by $\Omega=-\mu_\phi$. The range of the color spectrum is  (a,b) $[0,5\times 10^{-6}]$, (c,d) $[0,5\times 10^{-5}]$  (e,f) $[0,10^{-4}]$ for $h= 2\times 10^3 k_F$.}\label{KamikadoMolecule} \end{figure}

\textbf{Molecular spectral function.---} In \cref{KamikadoMolecule}, the molecular spectral function is shown for different interaction strengths $\epsilon_B/\epsilon_F$. Here the subfigures (a), (c) and (e) in the left column are obtained in the three-body truncation while (b), (d) and (f) in the right column result from the two-body truncation. It can be seen that a general feature of this spectral function is spectral weight that appears above a parabola centered around $p=k_F$ and that is defined by the frequency $\Omega_p=(p-k_F)^2 + m_{\phi,k=0}/A_{\phi,k=0}$ as derived in \cref{app:analyticalstructure}. The quasiparticle peak of the molecule follows a distorted dispersion relation which, in the strong-binding limit, tends to a free-molecule dispersion relation. Dependent on the interaction strength, at low momenta the molecular quasiparticle peak  lies outside of the particle-particle continuum and joins the continuum at finite momenta just to leave it again at higher momenta. More specifically, at low $\epsilon_B/\epsilon_F$, the  quasiparticle peak joins the continuum at a low momentum, which increases with interaction strength $\epsilon_B/\epsilon_F$. Likewise, the momentum at which the peak leaves the continuum again increases with $\epsilon_B/\epsilon_F$ as well.

Similar to a non-self-consistent $T$-matrix resummation, in our approach the molecular quasiparticle peak has a vanishing width when it is not embedded in the continuum. This can be seen analytically by inspecting the flow equation of the two-point function $G^R_{t}$ (see \cref{app:analyticalstructure} for details). Apart from the structure originating from the parabola-shaped particle-particle continuum and the quasiparticle peak, further structure exists within the parabola that originates from contributions in the RG flow where the Feynman diagrams are evaluated close to their poles (see \cref{app:analyticalstructure}).

The minimal energy of the parabola $\Omega_p$ is equal to the renormalized energy gap of the polaron, indicating a close relationship between the polaron at $p=0$ and the molecule at $p=k_F$, supporting the argument that both of these states overlap with the \emph{actual} groundstate of the system and possibly  with each other \cite{Bruun2010}. This finding can also be understood conceptually in a mean-field picture  where a bosonic minority particle at $p=0$ along with a majority fermions at the Fermi surface can be interpreted as either a polaron at $p=0$ or a molecule at $p=k_F$ (previously noted by Cui \cite{cui2020fermi}). Note that, because the particle-particle continuum in the molecular spectrum is shifted due to the renormalization of the boson gap, this effect can not be captured in a non-self-consistent approximation such as employed in Ref.~\cite{Schmidt2012}. In such an approximation spectral peaks distinct from the continuum  are present, that in our implementation are a part of the continuum.

Within the spectral functions obtained using $\Gamma_{2,k}$, the quasiparticle peak at $p=0$ ---located at approximately $m_{t,k=0}/A_{t,k=0}$---  is always at a finite energy whereas using $\Gamma_{3,k}$ it is moved closer to $\Omega=0$ and eventually attains $\Omega=0$ past the polaron-to-molecule transition. At the same time, the minimum of the parabola, given by $m_{\phi,k=0}/A_{\phi,k=0}$, detaches from $\Omega=0$ as the polaron is no longer the ground state. Hence using $\Gamma_{3,k}$ the effective mass (see Table~\ref{effectivemass}) of the molecule, which is negative at small $\epsilon_B/\epsilon_F$, diverges with increasing $\epsilon_B/\epsilon_F$ and eventually becomes positive at an interaction strength before the polaron-to-molecule transition. Beyond the transition the molecule is gapless at $p=0$ and its effective mass is positive. Using $\Gamma_{2,k}$, increasing $\epsilon_B/\epsilon_F$ makes the molecule dispersion flatter leading to an increasingly negative effective mass.

\section{Conclusion}
\label{sect:conclusion}

We investigated the phase diagram of strongly coupled Bose-Fermi mixtures in two dimensions. In order to make progress in the exploration of this complex phase diagram it is important to establish limits that can be understood controllably. To this end we focused  on the regime of fermion-dominated population-imbalance which, in the extreme imbalance limit, connects to the Fermi polaron problem where a single bosonic impurity interacts with a Fermi sea.
The opposite limit of a fermionic impurity coupled to a Bose-Einstein condensate corresponds to the Bose polaron problem which features qualitatively different physics. Already this asymmetry reflects the impact the interplay of different particle statistics has on the phase diagram away from the extreme population imbalanced limits.

In order to approach the problem we employed a functional-renormalization-group approach that allows to systematically incorporate high-order correlation functions. This enables us to reproduce the polaron-to-molecule transition in the single-boson limit which is a necessary condition for any theoretical approach that aims to describe this strong-coupling phase diagram. In contrast to the simpler three-dimensional case \cite{Chevy2006,Schmidt2011,Punk2009}, we showed that  three-body correlations have to be included  to describe the polaron-to-molecule transition in two dimensions and we obtain excellent agreement with ab-initio approaches \cite{Kroiss2014} that can be applied in the quantum impurity limit. 

Using the fRG we extended the analysis  to finite boson densities.  There, depending on the boson and fermion densities (or equivalently their  chemical potentials), we observed two phases:  a fermionic liquid with two Fermi seas in which all bosons are bound into molecules, themselves immersed in a majority Fermi sea, and a hybridized liquid in which the condensation of bosons leads to a mixing of the fermionic and molecular sectors \cite{Powell2005}. 

This hybridization and the associated mixing are not a result of the Hubbard-Stratonovich field used in our two-channel model, but they occur equally in atomic single-channel models whenever scattering vertices between fermions and bosons develop a pole in presence of a  boson condensate. In this regard, the phase diagram  away from the molecular phase at  $n_B\ll n_F$ shares a remarkable similarity to the  Bose polaron problem that describes the opposite limit of few fermions immersed in a Bose condensate, where the same hybridization mechanism leads to a \textit{crossover} between the polaron and molecule instead of a transition \cite{rath2013,yan2020bose}. 

 Naively, one may suspect that in a mixture of bosons and fermions as many particles as possible are bound into fermionic bound states in order to maximize attractive potential energy. This, however, does not take into account the properties of the system in two ways.  First, this argument neglects the fermionic nature of the bound states which leads to the formation of a molecular Fermi energy, representing a kinetic energy cost. As a result, when the bosonic density is increased, the molecular Fermi energy eventually exceeds the energy of the lowest-lying polaron state and the system enters the mixed phase.  

Second, the argument misses the fact that already in the limit of a vanishingly small boson density the formation of a bound state competes with the formation of a polaron state in which a single boson interacts collectively  with a large number of surrounding fermionic bath particles \cite{Imamoglu2021}. For a fixed interaction strength, the polaron state can thus profit more efficiently from an increased density of bath particles. Vice versa,  as the bath density is lowered polaron dressing looses efficiency so that  eventually the composite bound state becomes the new ground state (in absence of Coulomb interactions).

While the fRG approach employed in this work provides nontrivial insights into the phase diagram of the Fermi-Bose mixture, the approximations used are  insufficient to explore the phase diagram in its whole richness. First hints to a plethora of exciting phenomena can  already be inferred from  numerous quasiparticle features of the single-particle spectral functions uncovered using the FMR scheme in \cref{sect:Kamikado}, ranging from non-trivial effective mass renormalization and the  non-monotonous dispersion of  molecules, to incoherent parts in the spectral function reflecting quasiparticle instability. 

Indeed, for a more accurate description of such features it would be necessary to go beyond the gradient expansion we impose on our \emph{Ansatz} and instead allowing for an arbitrary momentum and frequency dependence of vertex functions. While such a treatment has been used in the three-dimensional case~\cite{Schmidt2011}, it remains challenging to implement numerically. Preliminary results \cite{jvm2021}, however, suggest that   including the full momentum- and frequency-dependence indeed cures the spurious non-monotonous behaviour of the polaron energy discussed in \cref{sect:polaron}. 
Far from being only of quantitative importance, such a fully momentum- and frequency-resolved approach could give new qualitative insight into the phase diagram, e.g. by allowing for the description of transitions to non-trivial molecular Fermi surface topology \cite{Sachdev2018} akin to Fulde-Ferrell-Larkin-Ovchinnikov phases in BCS superconductors \cite{Fulde1964,larkin1965}.

We did not  include Bose-Einstein condensation in our formalism. Its explicit inclusion  would  allow for the study of subregions of the mixed phase in which a bosonic condensate is accompanied by molecular or  fermionic Fermi seas. Additionally, the presence of a condensate will require the incorporation of a repulsive Bose-Bose interaction to ensure the mechanical stability of the condensate. Since the bosons are  strongly coupled to the fermions a strong renormalization of the boson-boson interaction has to be expected which may enhance or suppress the stability of Bose-Einstein condensation. While fermionic self-energy corrections are expected to play a subdominant role in the limit of strong population imbalance $n_F\gg n_B$, for a study of the phase diagram away from this limit these also become  an essential ingredient and may lead to  striking effects such as boson-mediated $p$-wave pairing at sufficient interaction strength \cite{Enss2009}. 

The question as to which vertices (i.e. correlation functions) to include in more refined approximations of our fRG scheme is dependent on the type of phases one may expect to govern the Bose-Fermi mixture phase diagram away from the strongly-imbalanced limit ---see the introductory \cref{SchematicPDintro}. Quite generally, and similar to variational techniques, in field theoretical approaches the range of phases one can discern is limited by the variety of ---potentially competing--- channels  taken into account in the renormalization procedure. In this regard, the strongly coupled Bose-Fermi mixtures  present a vast testbed to develop  comprehensive theoretical approaches to competing order where a manifold of scenarios and phases may unfold, including: phase separation between the fermionic species in case of repulsive effective interactions, competing bipolaron and trion formation, boson-mediated $s$- or $p$-wave pairing of fermions, fermion-induced phonon softening that may result in supersolidity,   higher-order pairing mechanisms such as  boson-mediated Cooper binding of trions and phases of Efimov-type states that  may condense depending on their statistics.

Moreover, as discussed in \cref{BBFcoupling}, the formation of bound states containing several bosons may be considered. However, in  ultracold quantum gases these higher-body bound states are usually subject to rapid decay to deeply bound states. The competition between such dissipative multi-particle losses and the formation of many-body phases is an intriguing perspective for future studies, posing a significant theoretical challenge that requires extension beyond equilibrium theory. 

Another compelling question is what the impact of  Coulomb interactions between the fermionic degrees of freedom may be. These long-range interactions will ultimately impose  limits on the universal connection between strongly coupled Bose-Fermi mixtures in atomically thin semiconductors and ultracold atoms (see \cref{tab:1}). Coulomb interaction can be expected to play  a key role in particular at low doping where screening becomes increasingly ineffective. Taking Coulomb interactions into account  may indeed suppress the formation of well-defined electronic and molecular Fermi surfaces and instead lead to qualitatively different physics even in the limit of extreme population imbalance $\epsilon_B/\epsilon_F$, where understanding the interplay of Coulomb interaction, favoring Wigner crystallization, and boson-mediated Fermi-Fermi interactions, remains an open challenge.

Considering the myriad of open questions, the full exploration of the phase diagram of two-dimensional Bose-Fermi mixtures  remains a formidable task. Due to the strong-coupling nature of the problem, uncovering the possible in- and out-of-equilibrium phases and phenomena will ultimately require a concerted effort between theory and experiment. Starting from  limiting cases, such as considered in this work, that can be controllably understood and combining \emph{ab initio} approaches with experimental observations will be key to tackle this outstanding challenge and can lead to new insight into effective descriptions of strongly coupled many-body quantum systems. 

\begin{acknowledgments}
We thank Ata\c c $\dot{\mathrm{I}}$mamo$\breve{\mathrm{g}}$lu, Eugene Demler and Wilhelm Zwerger for interesting discussions and valuable input. 
R.~S. and F.~R. are supported by the Deutsche Forschungsgemeinschaft (DFG, German Research Foundation) under Germany's Excellence Strategy -- EXC-2111 -- 390814868. J. v. M. is supported by a fellowship of the International Max Planck Research School for Quantum Science and Technology (IMPRS-QST).

\end{acknowledgments}

\appendix
\section{Flow equations}\label{sect:AppFloweqs}
\subsection{Vertex projections}
\label{sect:AppProj}
The  $n$-point functions considered in this work are obtained from the effective flowing action $\Gamma_k$ using the following projections: 
\begin{align}
G^{-1}_{\psi,k}(\pv,\omega)&= \frac{\delta}{\delta \psi(\pv,\omega)} \frac{\delta}{\delta \psi^{\ast}(\pv,\omega)} \Gamma_k \ ,\nonumber\\
G^{-1}_{\phi,k}(\pv,\omega)&= \frac{\delta}{\delta \phi(\pv,\omega)} \frac{\delta}{\delta \phi^{\ast}(\pv,\omega)} \Gamma_k \ , \nonumber \\
G^{-1}_{t,k}(\pv,\omega)&=\frac{\delta}{\delta t(\pv,\omega)} \frac{\delta}{\delta t^{\ast}(\pv,\omega)} \Gamma_k  \ , \nonumber \\
\frac{h_k}{(2\pi)^{3/2}}&= \frac{\delta}{\delta t(\zerov,0)} \frac{\delta}{\delta \phi^{\ast}(\zerov,0)} \frac{\delta}{\delta \psi^{\ast} (\zerov,0)} \Gamma_k  \ , \nonumber\\
\frac{\lambda_k}{(2\pi)^3}&=\frac{\delta}{\delta \psi (\zerov,0)} \frac{\delta}{\delta t(\zerov,0)} \frac{\delta}{\delta t^{\ast}(\zerov,0)} \frac{\delta}{\delta \psi^{\ast}(\zerov,0)} \Gamma_k \ .
\end{align}

\subsection{Gradient expansion parameters}
\label{sect:gradexppar}
The parameters of the gradient expansion  of the two-point functions $G_{\phi,\psi,t,k}^{-1}$ are given by 
\begin{align}
m_\psi&= G^{-1}_{\psi,k}(\pv,\omega)|_{\pv^2= \epsilon_F,\omega=0} \ , \nonumber\\
A_\psi&=\partial_{-i\omega} G^{-1}_{\psi,k}(\pv, \omega)|_{\pv^2= \epsilon_F,\omega=0} \ , \nonumber\\
m_\phi &= G^{-1}_{\phi,k}(\pv,\omega)|_{\pv=\zerov,\omega=0} \ , \nonumber\\
A_\phi&= \partial_{-i\omega} G^{-1}_{\phi,k}(\pv,\omega)|_{\pv=\zerov,\omega=0}  \ , \nonumber\\
m_t &= G^{-1}_{t,k}(\pv,\omega)|_{\pv=\zerov,\omega=0} \ , \nonumber\\
A_t &= \partial_{-i\omega} G^{-1}_t(\pv,\omega)|_{\pv=\zerov,\omega=0} \ . \quad 
\end{align}
This prescription can then be used in \cref{Gphiflow,Gpsiflow,Gtflow,hflow,lambdaflow} to determine the flow of the parameters. In \cref{sect:AppExplicitFlowEqs} we provide the explicit form of the flow equations.

\section{Initial conditions of the flow}
\label{app:initial}
The initial conditions for the flow are obtained by setting $\Gamma_{k=\Lambda}=S + const$. This implies that $m_{\psi,k=\Lambda}= 0$, $m_{\phi,k=\Lambda}= - \mu_\phi$, $m_{t,k=\Lambda}= m_t$ and   $A_{\psi,k=\Lambda}= A_{\phi,k=\Lambda}=A_{t,k=\Lambda}= 1$. The initial conditions for the interaction vertices are given by $\lambda_{k=\Lambda}=0$ and $h_{k=\Lambda}=h$. 

The initial condition for the detuning of the molecule $m_t$ is obtained from the physical renormalization condition that in the two-body problem a bound state of energy $\epsilon_B$ forms between the boson and the fermion species. Within an fRG approach, this condition is ensured by, first, setting $\mu_{\psi}$ and $\mu_{\phi}$ to sufficiently negative values so that either species has a zero density. Furthermore, we set $\mu_{\psi}+\mu_{\phi}=-\epsilon_B$. This ensures that the energetic cost to bring up a particle of both species from vacuum is given by the binding energy. Finally, the condition for the bound-state formation is given by $m_{t,k=0}=0$ and $m_{t,k}\geq 0 $. This condition guarantees that the chemical potentials are tuned correctly to the boundary between the vacuum state and the state comprised of a molecule submersed in vacuum.

The flow of the three-body vertex does not have to be taken into account, since $\lambda_k$ does not feed back into the solution of the two-body problem. Similarly, in the two-body problem the flow equations of  $G_{\psi,k}^{-1}(P)$ and $G_{\phi,k}^{-1}(P)$ evaluate to zero because the poles of their propagators in \cref{Gphiflow,Gpsiflow} lie in the same half of the complex plane and thus their frequency contour integrals  evaluate to zero. Physically, this is because neither of the particle species has a finite density which would be required to generate a renormalization of the particle self-energies by particle-hole fluctuations. The molecule on the other hand is renormalized by a particle-particle diagram and thus does not require a finite density of bosons or fermions. 

As the three body-vertex is not relevant in the two-body problem, the Yukawa term $h$ does not renormalize. After evaluation and projection of \cref{Gtflow}, the flow equations of $\mphi$ and $\Aphi$ are therefore given by 
\begin{align}
\partial_k \mphi&= \frac{h^2 k }{2 \pi } \frac{1 }{2 k^2 - \mu_\phi - \mu_\psi} \  \label{vacuumpartialphi}
\end{align}
and 
\begin{align}
\partial_k \Aphi&= -\frac{h^2 k }{2 \pi } \frac{1 }{(2 k^2 - \mu_\phi - \mu_\psi)^2} \ . \label{vacuumpartialAphi}
\end{align}
 Using $m_{t,k=\Lambda}= m_{t,k=0}+ \int_0^\Lambda dk( \partial_k m_{t,k})$ and $m_{t,k=0}=0$ this reproduces \cref{mphiinitial}. Note that since in this few-body calculation no Fermi surfaces are present,  the regulators are proportional to $\Theta(|\pv|-k)$ for all particles involved. 

\section{The polaron energy within the gradient expansion scheme}\label{app:abovebelow}
\begin{figure}
	\begin{center} 
		\includegraphics[width=.99\linewidth]{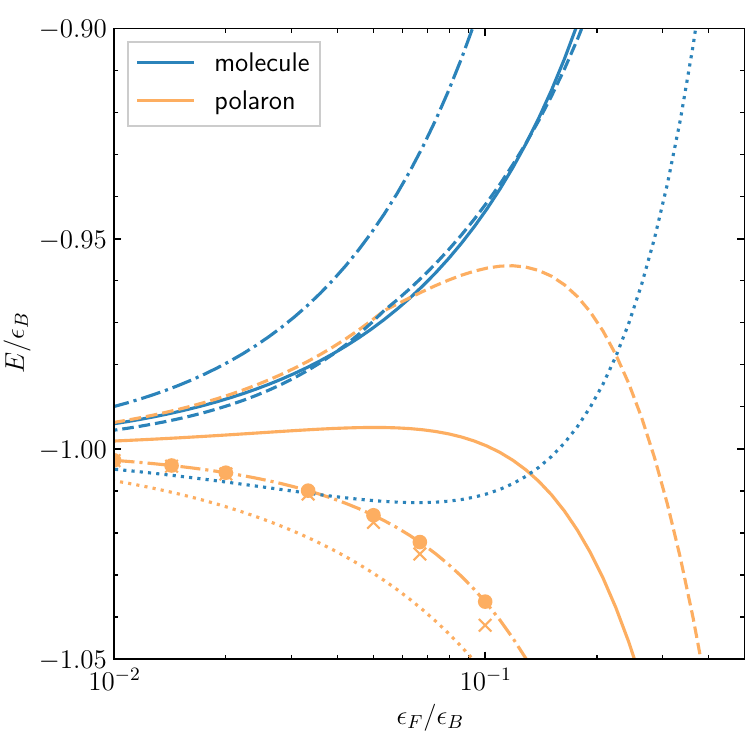} 	\end{center}
	\caption{Energies of the polaron (orange lines) and the molecule (blue lines) as a function of dimensionless interaction strength using different fRG implementations. The solid lines show energies using $\Gamma_{2,k}$ while the dash-dotted lines show a two-step implementation of $\Gamma_{2,k}$ in which the molecule is renormalized in the first step and the $\phi$-boson is renormalized in the second step. The dashed lines show the $\Gamma_{3,k}$ implementation and the dotted lines show a $\Gamma_{3,k}$ implementation in which the flow of $h_k$ is neglected by setting $\partial_k h_k=0$. The cross markers show the polaron energies resulting from a non-self-consistent $T$-matrix approximation  \cite{Schmidt2012}, while the dot markers show the result of this calculation in a gradient expansion.}\label{abovebeloweB} \end{figure}
	
In this appendix we discuss the weak non-monotonous behavior of the polaron energy as a function of $\epsilon_F/\epsilon_B$. Generally, the polaron energy lies approximately within a range of $\pm \epsilon_F$ around the value of $-\epsilon_B$. For small values of $\epsilon_F/\epsilon_B$ it is thus not surprising to see that $E\to -\epsilon_B$. Previous calculations \cite{Schmidt2012,Zoellner2011, Parish2011, Parish2013,Kroiss2014,Vlietinck2014}  indicate that for all values of $\epsilon_F/\epsilon_B$ the value of the polaron energy should lie below $-\epsilon_B$, in disagreement with the results shown in \cref{pmenerg}. This discrepancy highlights one of the major shortcomings of fRG, namely the dependence on regulators and on the truncation scheme. 

To analyze this finding in detail in \cref{abovebeloweB} we show the polaron and the molecule energies using different truncation and regulator schemes. As one can see, the $\Gamma_{2,k}$ truncation (solid line) presented also in \cref{pmenerg} results in polaron energies above $-\epsilon_B$. If, however, the same truncation is used and the regulators are changed such that the renormalization group flow consists of two steps, where in the first step only the molecule and in the second step only the minority particle is allowed to  flow, this results (dash-dotted) in polaron energies strictly below $-\epsilon_B$. Within this scheme, however, the resulting molecule energy lies higher than before. Effectively, by treating the molecule and polaron on different footing (i.e. by treating them in different steps of the fRG) we have improved the polaron energy at the cost of a higher-lying molecular energy. Interestingly, this two-step calculation is closely related to the results obtained within the variational approach in Ref.~\cite{Zoellner2011} and the ladder resummations performed in Ref.~\cite{Schmidt2012} (crosses). If the full frequency- and momentum-resolved $T$-matrix in these two approaches is replaced by a gradient expansion of the $T$-matrix, the resulting method is equivalent to the two-step fRG. The results for this modified variational/diagrammatic calculation are shown as dots and coincide with the two-step calculation as expected. A similar equivalence of the FMR scheme is discussed in \cref{App:Equivalence}.

Within the $\Gamma_{3,k}$ calculation presented in \cref{abovebeloweB} (dashed) and also in Fig.~\ref{pmenerg} the polaron energy lies again above $-\epsilon_B$ for small $\epsilon_F/\epsilon_B$. If, however, the flow of $h_k$ is turned off  (dotted) the polaron energy lies once again strictly below $-\epsilon_B$. These observations illustrate the dependence of the absolute values of the energy on the regulators and   truncation employed. For example, the flow of the Yukawa vertex $h_k$ has a significant impact on the polaron energy, which is likely due to its point-like projection.

Although the relative deviations of these energies are only of the order of a few percent, we do not expect that the used fRG schemes are a reliable method of determining the \textit{absolute} energy of the polaron and the molecule. Most of the variational approaches, however, do not consider the polaron and the molecule on an equal footing and therefore can produce ambiguous results when one considers transitions which depend on relative energy differences between the emergent quasiparticles.
We thus believe that, by treating the polaron and the molecule on equal footing within a unified renormalization approach, the fRG scheme captures the qualitative physics correctly  and can thus make qualitative predictions about transition in the quantum many-body system. 

\section{Three-body vertex in the vacuum three-body limit}
\label{threebodyvacuum}
To determine the value of the three-body vertex in the limit where two $\psi$-particles and a single $\phi$-particle are present, we solve the flow equations under the initial conditions of the two-body problem discussed in \cref{app:initial}, and additionally take into account the flow of $\lambda_k$. Since $\lambda_k$ corresponds to the \textit{on-mass-shell} scattering of a molecule and a quasi-free excess fermion, we supplement the two-body initial conditions by setting the fermionic chemical potential to a small, negative value $\mu_\psi=0^-$ while we set $\mu_\phi=-\epsilon_B- 0^-$.

As in the two-body case, $G_{\phi,k}^{-1}$ and $G_{\psi,k}^{-1}$ do not flow and as a result neither does $h_k$. Consequentially, the flow of $m_{t,k}$ and $A_{t,k}$ is not influenced by the flow of $\lambda_k$ such that according to \cref{vacuumpartialphi,vacuumpartialAphi}
\begin{align}
 m_{t,k}=\frac{h^2}{8 \pi} \log \left(1 + \frac{2 k^2}{\epsilon_B}\right)
\end{align}
and similarly
\begin{align}
    A_{t,k}= 1+ \frac{h^2}{8 \pi} \left( \frac{1}{\epsilon_B+ 2 k^2}- \frac{1}{\epsilon_B+ 2 \Lambda^2}  \right)\ . \label{app:threebodyvacuumAt}
\end{align}
The flow equation of $\lambda_k$ is then given by 
\begin{align}
\partial_k\lambda_k= \frac{k (h^2 + 2\lambda_k k^2 +\lambda_k \epsilon_B )^2}{\pi A_{t,k}(2 k^2 + \epsilon_B)^2 (3 k^2 + 2 m_{t,k}/A_{t,k}-2 \mu_\psi)} \label{app:threebodyvacuummt}
\end{align}
leading to $\lambda_{k=0}=-h^2 /\epsilon_B$ for large values of $h$.

\section{Bose-Bose-Fermi coupling in the three-body limit and at finite density}\label{BBFcoupling}
The truncations considered in the main text neglect the emergence of a Bose-Bose-Fermi (BBF) coupling (and other higher-order couplings). In this Appendix we seek to explore the relevance of this coupling. From a physical standpoint, unlike the Fermi-Fermi-Bose (FFB) coupling  $\lambda_k$, the BBF coupling does not suffer from Pauli blocking and may thus be considerably stronger, potentially resulting in the formation of bound states containing more than one boson.  

\begin{figure*}[t]
  \begin{center}
		\includegraphics[width=\linewidth]{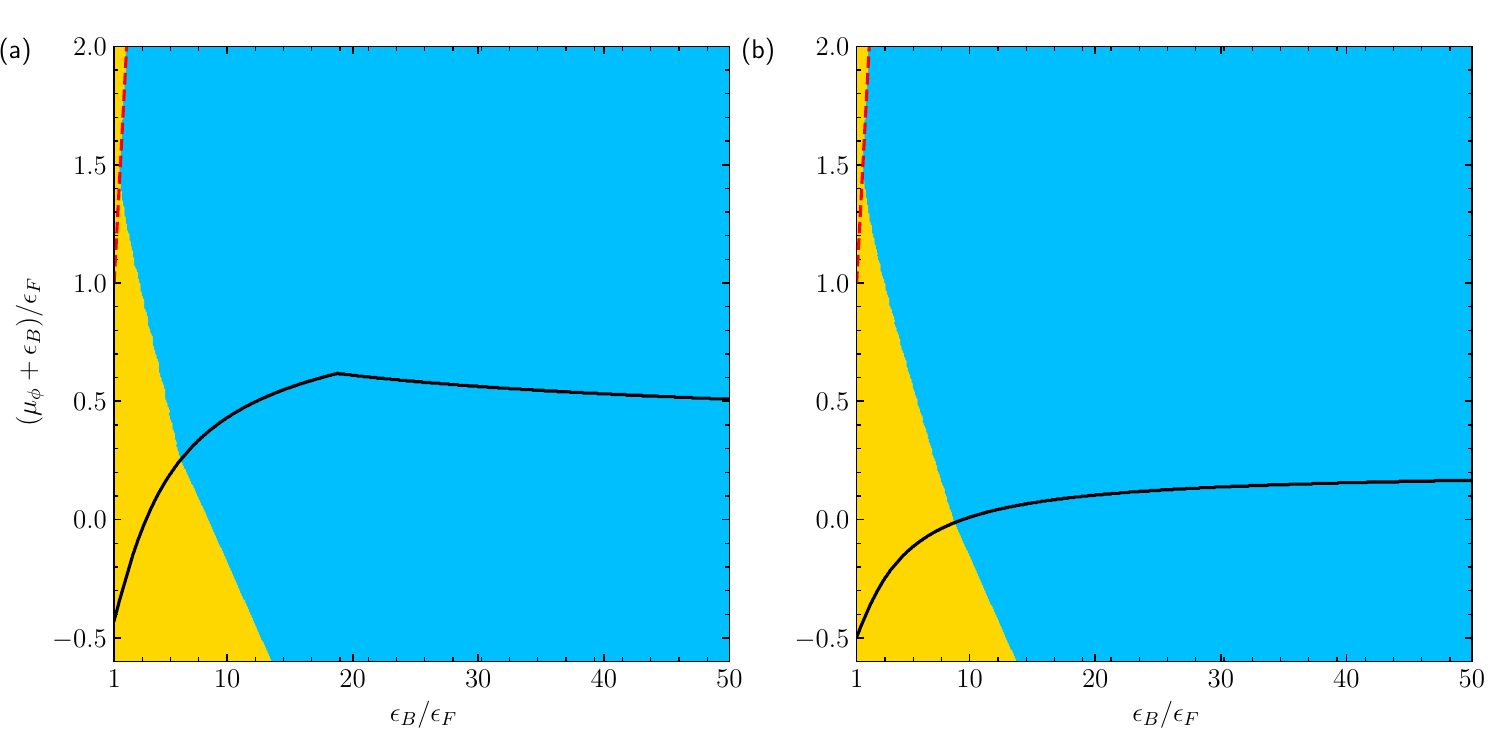}
  \end{center}
	\caption{Instability towards higher-order bound-state formation. Indication of a divergence of $\tau_k$ for different ratios of $\mu_{\phi}/\epsilon_F$ and $\epsilon_B/\epsilon_F$ using $\Gamma_{3,k}$ (a) and $\Gamma_{2,k}$ (b). The blue (dark gray) regions indicate the divergence of $\tau_k$ at finite $k$ (thus resulting in a potential instability towards the formation of a Boson-Boson-Fermion bound state), while the yellow (light gray) regions indicate that $\tau_k$ had not diverged when the flow ended as discussed in \cref{sect:finitedens}. The solid black line shows the transition to the finite boson density regime, while  the dashed red line indicates the path along which $\mu_\phi=0$, as in \cref{chemicalpot}.}\label{BBFdivergence} \end{figure*}

First, we study the BBF coupling in the limit where two $\phi$ bosons and a single $\psi$ fermion are present. We define the corresponding coupling vertex $\tau_k$ as 
\begin{align}
    \tau_k \int_x \phi^*_x t^*_x t^{\phantom{*}}_x \phi^{\phantom{*}}_x \ . 
\end{align}
The flow equations given in \cref{Gphiflow,Gpsiflow,Gtflow,hflow}  (excluding the flow of $\lambda_k$) are then complemented by the RG flow of $\tau_k$
\begin{align}
\partial_k \tau_k &=  -\tau_k^2\tilde{\partial}_k \int_Q G^c_{t,k}(Q)\left[G^c_{\phi,k}(Q)+ G^c_{\phi,k}(-Q)\right]\nonumber\\
&-h_k^4\tilde{\partial}_k \int_Q G^c_{t,k}(Q)G^c_{\psi,k}(Q)^2 G^c_{\phi,k}(-Q)\nonumber\\
&+2 h_k^2 \tau_k \tilde{\partial}_k \int_Q G^c_{t,k}(Q)G^c_{\psi,k}(Q) G^c_{\phi,k}(-Q)\ . \label{lambdaflow}
\end{align}
These equations are solved using the initial conditions of the two-body problem discussed in \cref{app:initial} where, unlike in \cref{threebodyvacuum}, we consider the \textit{on-mass-shell} scattering of a molecule and a quasi-free excess boson such that $\mu_\phi=0^-$ and $\mu_\psi=- \epsilon_B - 0^-$. As in the previous section $G_{\phi,k}^{-1}$, $G_{\psi,k}^{-1}$ and $h_k$ do not flow such that $m_{t,k}$ and $A_{t,k}$ are given by \cref{app:threebodyvacuumAt,app:threebodyvacuummt}, respectively. We find that  $\tau_k$ flows from $\tau_{k=\Lambda}=0$ to negative values before diverging at $k>0$ and continuing to flow to $\tau_{k=0}=h^2 /\epsilon_B$ at the end of the flow. The divergence indeed indicates the formation of three-body bound states in the vacuum limit as predicted in Refs.~\cite{Pricoupenko2010,Levinsen2014c,Naidon2017}. Our results show that these can, in principle, be captured using our fRG technique. We now demonstrate that this treatment can be extended to finite density. 

To this end, we study the behavior of the BBF coupling at finite density. Thus, we apply the initial conditions used for \cref{chemicalpot} by tuning the binding energy and the boson chemical potential at a fixed Fermi energy. In order to simplify the calculation, however, here we do not choose a fully self-consistent calculation, but rather treat the BBF coupling as an observing flow that does not feed back into the renormalization of the other coupling constants. Hence, the flow of $\tau_k$ is influenced by the flow of $\lambda_k$ (but not vice versa). In this framework at finite density the RG flow of $\tau_k$ picks up another term given by: 
\begin{align}
  - h_k^2 \lambda_k \tilde{\partial}_k \int_Q G^c_{t,k}(Q)G^c_{\psi,k}(Q)^2 \ . 
\end{align}

We now turn to the question under which conditions a divergence of $\tau_k$ occurs during the flow. The result of this calculation is shown in \cref{BBFdivergence}. As can be seen, in both truncations, $\Gamma_{2,k}$ and $\Gamma_{3,k}$, the coupling constant $\tau_k$ diverges  for  most of the  combinations of $\mu_\phi/\epsilon_B$ and $\epsilon_B$ shown in \cref{chemicalpot}. Only at weaker interaction strengths when the boson is gapped strongly does the coupling constant remain finite. This shows the importance of `non-Pauli-blocked' coupling channels such as $\tau_k$ which lead to bound states containing more than one boson, especially at a finite boson density where these are not suppressed.

\section{Contour integrals leading to the flowing effective potential}
\label{sect:effPot}
Expanding \cref{effpotflow1} we obtain that $\partial_k U_k(\sigma_{\text{eq}})$ has the following structure
\begin{align}
\partial_k U_k&(\sigma_{\text{eq}})\propto  \xi_\sigma \int_{P}\Theta_{\sigma,k}(\pv,\omega) \partial_k \left[\frac{1}{\Theta_{\sigma,k}(\pv,\omega)}-1\right]  \nonumber\\
&+  \xi_\sigma \int_{P}[1-\Theta_{\sigma,k}(\pv,\omega)]  G_{\sigma,k}(\pv, \omega) \partial_k G^{-1}_{\sigma,k}(\pv,\omega)\label{effpotsimplification} \ . 
\end{align}
The integrand in the first term in \cref{effpotsimplification}  does not have a pole in the frequency domain as the $\Theta_{\sigma,k}$-functions are frequency-independent. Stemming from the construction of the quantum field theory and the convergence factor of $e^{i \omega 0^+}$, this integral thus evaluates to zero. The second term, in contrast, possesses a pole within $G_{\sigma,k}$ and therefore does not vanish, yielding \cref{partialU}. Note that because the integrand only falls off fast enough due to the convergence factor, these $\omega$ contour integrals need to be closed within the upper half of the complex plane. Consequently, the second term in \cref{partialU}  always vanishes. In order to yield a finite value it would  require the polaron to develop a finite density which we do not allow for within our phase identification scheme.

\section{Frequency- and momentum-resolved spectral function}\label{sect:AppImag}
Here we provide the explicit flow equations used to obtain the frequency- and momentum-resolved spectral functions in \cref{sect:Kamikado}. Furthermore we comment on the analytical structure of these equations and how it relates to the structure of the particle-hole/particle-particle continua visible in \cref{KamikadoPolaron,KamikadoMolecule} and the lifetimes of the molecule and the polaron. Finally, we show the close correspondence of this method to  $T$-matrix approximation schemes. 

\subsection{Frequency- and momentum-resolved flow equations}
\label{Kamikadoexplicit}

In order to compute the frequency- and momentum-resolved spectral functions within the FMR scheme, for a given value of $\epsilon_F$, $\mu_{\phi}$ and $\epsilon_B$ in a first step
the flow of the expansion parameters is computed as detailed in \cref{sect:polaron} and  \cref{sect:gradexppar}. In a second step the solutions of the flow equations for the different gradient expansion parameters are plugged into the rhs. of the flow equations given in \cref{Gphiflow,Gpsiflow,Gtflow,hflow,lambdaflow}. This time, however, the flow equations are considered for arbitrary external momentum and frequency. Next, the Matsubara integration is performed as usual and the complex frequency $\omega$ of $P=(\pv,\omega)$  is continued to the real frequency axis $i \omega \to \Omega + i 0^+$. For every external  frequency and momentum the flow equations of the retarded Green's function can then be computed and yield

\begin{widetext}
\begin{align}
\partial_k \left(G^{R}_{\phi,k}\right)^{-1}(\pv,\Omega)&= h_k^2k  \int \frac{ d\theta }{(2\pi)^2}\Bigg[\frac{1}{\Aphi}\frac{\Theta\left(  \epsilon_F -\pv^2 - 2|\pv|k \cos(\theta)- 2k^2\right)\Theta\left(  \epsilon_F-\pv^2-k^2 -2|\pv|k\cos(\theta)\right)}{- k^2 /2 - \pv^2 + \epsilon_F - 2 |\pv|k \cos(\theta) - \Omega + \mphi/\Aphi - i 0^+} \nnl
&+\frac{1}{\Aphi}\frac{\Theta\left(\pv^2 +\epsilon_F +2|\pv|\sqrt{\epsilon_F-k^2} \cos(\theta)- 2k^2\right)\Theta\left(\epsilon_F - k^2\right)}{k^2/2 + \pv^2/2 + \epsilon_F/2 +|\pv| \sqrt{\epsilon_F -k^2}\cos(\theta)- \Omega + \mphi/\Aphi- i 0^+}\Bigg]\label{KamikadoFlowDown}\\
\partial_k \left(G^{R}_{t,k}\right)^{-1}(\pv,\Omega)&=-\frac{k}{2\pi} \lambda_k  \Theta(\epsilon_F- k^2 ) + h_k^2 k \int \frac{ d\theta }{(2\pi)^2} \Bigg[\frac{1}{\Ad} \frac{\Theta(\pv^2  + 2|\pv|k \cos(\theta)-\epsilon_F)}{2 k^2 + \pv^2 -\epsilon_F- \Omega+2 |\pv|k \cos(\theta) + \md/\Ad - i0^+ 
}\nnl
&+ \frac{1}{\Ad}\frac{\Theta(\pv^2 + \epsilon_F + 2|\pv| \sqrt{\epsilon_F+k^2 }\cos(\theta))}{2 k^2 + \pv^2 + \epsilon_F + 2 |\pv| \sqrt{\epsilon_F + k^2 }\cos(\theta) - \Omega + \md/\Ad - i0^+}\Bigg]\label{KamikadoFlowMol}\ . 
\end{align}
\end{widetext}

\subsection{Equivalence to a non-self-consistent \texorpdfstring{$T$}{T}-matrix resummation}
\label{App:Equivalence}
In this subsection we show the close correspondence between the FMR scheme (\cref{sect:Kamikado}) and diagrammatic ladder approximations. More specifically, we show that a non-self-consistent implementation of the FMR method exactly corresponds to the result obtained for the molecule in non-self-consistent $T$-matrix resummation as presented in Ref.~\cite{Schmidt2012}.  

Using only bare quantities on the rhs. of the flow equation and performing the frequency integration in the quantum impurity limit, the flow of the retarded inverse molecule propagator reads
\begin{align}
  \partial_k& \left(G^{R}_{t,k}\right)^{-1}(\pv,\Omega)=\nonumber \\ & -  h_k^2 \partial_k \int\frac{d\qv}{(2 \pi)^2} \frac{\Theta((\pv -\qv)^2- k^2)\Theta(\qv^2-\epsilon_F-k^2)}{ \qv^2 - \epsilon_F + (\pv -\qv)^2 - \mu_\phi - \Omega - i0^+ }\ .
\end{align}
Here we used that within this approximation $\lambda_k=0, h_k=h, A_{\phi,k}=1 , m_{\phi,k}= - \mu_\phi, A_{\psi,k}=1$ and $m_{\psi,k}= -\epsilon_F$. Note that, since we only use bare quantities on the rhs., we have  $\tilde{\partial}_k= \partial_k$. Thus we can  perform the $k$-integration analytically and obtain
\begin{widetext}
\begin{align}
   \left(G^{R}_{t,k=0}\right)^{-1}(\pv,\Omega)&=\left(G^{R}_{t,k=\Lambda} \right)^{-1}(\pv,\Omega) - h^2 \int_{\qv^2>\epsilon_F}\frac{d\qv}{(2 \pi)^2} \frac{1}{ \qv^2 - \epsilon_F + (\pv -\qv)^2 - \mu_\phi - \Omega - i0^+ }\nnl &+h^2  \int\frac{d\qv}{(2 \pi)^2} \frac{\Theta((\pv -\qv)^2- \Lambda^2)\Theta(\qv^2-\epsilon_F-\Lambda^2)}{ \qv^2 - \epsilon_F + (\pv -\qv)^2 - \mu_\phi - \Omega - i0^+ }\nnl
   & =h^2 \int \frac{d \qv}{(2 \pi)^2} \frac{1}{\epsilon_B + 2 \qv^2} - h^2 \int_{\qv^2>\epsilon_F}\frac{d\qv}{(2 \pi)^2} \frac{1}{ \qv^2 - \epsilon_F + (\pv -\qv)^2 - \mu_\phi - \Omega - i0^+ }\nnl 
   &+h^2  \int\frac{d\qv}{(2 \pi)^2}\left( \frac{\Theta((\pv -\qv)^2- \Lambda^2)\Theta(\qv^2-\epsilon_F-\Lambda^2)}{ \qv^2 - \epsilon_F + (\pv -\qv)^2 - \mu_\phi - \Omega - i0^+ }-\frac{\Theta(q-\Lambda)}{\epsilon_B + 2 \qv^2}\right)\nnl
  & \stackrel{(\Lambda\to \infty)}{=}- h^2\left( \frac{i\pi + \log \left( \frac{\epsilon_B}{\Omega + \epsilon_F +\mu_\phi - \pv^2 /2 + i0^+ }\right)}{8\pi}  - \int_{\qv^2<\epsilon_F}\frac{d\qv}{(2 \pi)^2} \frac{1}{ \qv^2 - \epsilon_F + (\pv -\qv)^2 - \mu_\phi - \Omega - i0^+ }\right) \label{MolPropNSCT} \ , 
\end{align}
\end{widetext}
reproducing the molecular results presented in Ref.~\cite{Schmidt2012}. Furthermore, similar analysis shows that performing a modified non-self-consistent two-step fRG of the FMR scheme also reproduces the polaron results presented in Ref.~\cite{Schmidt2012}. In such a two-step approach the molecular propagator is renormalized in the first step as described in \cref{MolPropNSCT}, and in the second step the minority propagator is renormalized as prescribed by \cref{Gphiflow}. In this second step, on the rhs. the coupling constants along with the majority propagator appear in their bare form and the molecular propagator with its full frequency- and momentum-dependence obtained in the first RG step is used instead of a gradient expansion. The polaron energy resulting from this calculation is shown as crosses in \cref{abovebeloweB}. It is worth noting, however, that as a starting point for the second step one may also perform a gradient expansion of the molecular propagator of the form 
\begin{widetext}
\begin{align}
    \left(G_{t}^{R,2^{\text{nd}} }\right)^{-1} (\pv,\Omega)= \left(G^{R}_{t,k=0}\right)^{-1}(0,0) - \left(-\Omega - i0^+  + \frac{\pv^2}{2}\right) \left[\partial_\Omega \left(G^{R}_{t,k=0}\right)^{-1}(0,\Omega)\right]_{\Omega=0}
\end{align}
\end{widetext}
and still obtain similar results (dash-dotted lines and dot markers in \cref{abovebeloweB}). This then directly corresponds to a version of the FMR scheme used to obtain spectral functions in which the renormalization of the molecule and the minority is divided into two consecutive steps while retaining the gradient expansion on the rhs. of the flow equations.

\subsection{Analytical structure of the FMR flow equations}\label{app:analyticalstructure}
In the following we analyze the analytical structure of the flow of the retarded inverse Green's function of the molecule and how it is reflected in the spectral functions shown in  \cref{KamikadoMolecule}. A similar analysis can be performed on the retarded inverse Green's function of the polaron as well. 

Within the FMR scheme of analytic continuation, the retarded self-energy can only pick up a non-vanishing imaginary part in the limit of $i 0^+\to 0 $ if during the flow one integrates over a pole caused by $i\omega \to \Omega$. In that case we have encountered a pole in the flow that is only avoided by the use of a retarded frequency and the self-energy picks up an imaginary part that is non-vanishing for all $i 0^+$.

Contributions to the spectral function defined in \cref{eq:defSpecFun} can have two different origins. Either the Green's function picks up an imaginary part in the course of the flow as described above, or the inverse Green's function tends to $i0^+$ resulting in a sharp excitation feature in the spectral function. In the former case the corresponding states are part of a particle-particle continuum of states with a finite lifetime, whereas in the latter case the corresponding excitations have an infinite lifetime.

\begin{figure}[b]
	\begin{center} 
		\includegraphics[width=\linewidth]{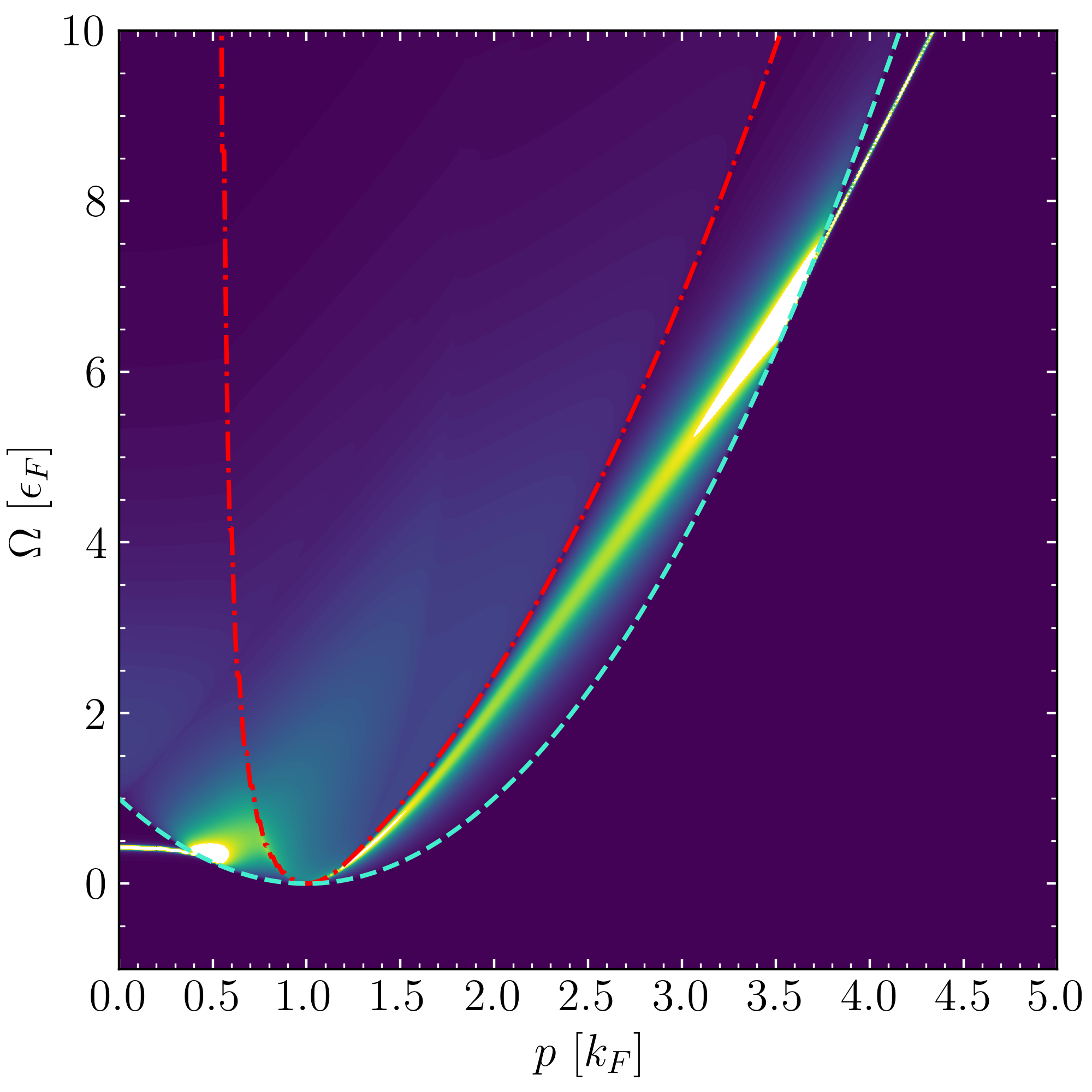} 	\end{center}
	\caption{Molecular spectral function at $\epsilon_B/\epsilon_F=20$ within $\Gamma_{2,k}$. The minimal frequencies $\Omega_{\text{min},1}= \text{min}\left[\Omega_{\text{min},1}^<,\Omega_{\text{min},1}^> \right]$ and $\Omega_{\text{min},2}$ are shown in red (dash-dotted) and mint-color (dashed), respectively.}\label{Kamikadolines} \end{figure}

Inspecting the second term of \cref{KamikadoFlowMol}, we see that it causes the self-energy to develop an imaginary part if during the flow
\begin{align}
    \Omega=2 k^2 + \pv^2 - \epsilon_F +2 |\pv| k\cos(\theta) +  \frac{\md}{\Ad}
\end{align}
while $\pv^2 + 2 |\pv| k \cos(\theta)-\epsilon_F>0$. For $\pv^2 <\epsilon_F$ the minimal frequency for which this can occur is given by 
\begin{align}
    \Omega^{<}_{\text{min},1}=\left[ \frac{\md}{\Ad} + 2 k^2 \right]_{k= \frac{\epsilon_F-\pv^2}{2 |\pv|} }\ ,
\end{align}
where we made use of the fact that $\md/\Ad$ decreases monotonically during the flow.
In turn, for $\pv^2>\epsilon_F$ this frequency is given by
\begin{align}
    \Omega^{>}_{\text{min},1}= \min_{k,0\leq k\leq \frac{\pv^2 - \epsilon_F }{2 |\pv| } } \pv^2 - 2|\pv|k -\epsilon_F + 2 k^2 + \frac{\md}{\Ad} \ . 
\end{align}
Analogously, the minimal frequency for which the third term of \cref{KamikadoFlowMol} leads to an imaginary part is given by 
\begin{align}
    \Omega_{\text{min},2}&= \min_{\substack{k,\ k>0, \\  k\leq \frac{|\epsilon_F-\pv^2|}{2|\pv|}}} k^2 + \left(|\pv|- \sqrt{\epsilon_F + k^2}\right)^2+\frac{ \md}{ \Ad}\ .
\end{align}
Numerically we find that this is solved by
\begin{align}
    \Omega_{\text{min},2}&=  \left(|\pv|- \sqrt{\epsilon_F }\right)^2+\left[\frac{ \md}{ \Ad}\right]_{k=0} 
\end{align}
for the interaction strengths studied here.

In \cref{Kamikadolines}, the spectral function from \cref{KamikadoMolecule}(f) is shown along with the minimal frequencies $\Omega_{\text{min},1}$ and $\Omega_{\text{min},2}$. As it can be seen, these frequencies determine the onset of the particle-particle continua. Furthermore, as the molecule peak at low and high momenta lies outside the boundaries of the continua, the corresponding excitations possess an infinite lifetime within this renormalization scheme.

\section{Explicit flow equations}\label{sect:AppExplicitFlowEqs}
In this appendix we provide the explicit flow equations of parameters of the  gradient expansion. These flows are  obtained as described in \cref{sect:AppFloweqs}, and for completeness we state them here explicitly. Note that we state the flow equations as used in \cref{sect:finitedens}. These are a generalization of the flow equations used in  \cref{sect:polaron} and  as such may also be used there.

\begin{widetext}
\subsection{Boson renormalization}
\begin{align}
\partial_k \md&= \frac{h_k^2 k}{\pi}\left[\Theta(\epsilon_F - 2 k^2)\left( \frac{\Theta\left(\epsilon_F- k^2 + 2 \frac{\mphi}{\Aphi}\right) }{2 \mphi + \Aphi(k^2 + \epsilon_F )}+ \frac{\Theta\left(k^2 + 2 \frac{\mphi}{\Aphi}\right)}{2\mphi - \Aphi(k^2 - 2\epsilon_F)}\right)+ \frac{\Theta\left(-2 \frac{\mphi}{\Aphi}- (k^2 + \epsilon_F)\right)}{-2\mphi + \Aphi (k^2 - \epsilon_F)}\right]\\
\partial_k \Ad&= \frac{2\Aphi h_k^2 k}{\pi}\left[-\Theta(\epsilon_F - 2 k^2)\left( \frac{\Theta\left(\epsilon_F- k^2 + 2 \frac{\mphi}{\Aphi}\right) }{[2 \mphi + \Aphi(k^2 + \epsilon_F )]^2}+ \frac{\Theta\left(k^2 + 2 \frac{\mphi}{\Aphi}\right)}{[2\mphi - \Aphi(k^2 - 2\epsilon_F)]^2}\right)+ \frac{\Theta\left(-2 \frac{\mphi}{\Aphi}- (k^2 + \epsilon_F)\right)}{[-2\mphi + \Aphi (k^2 - \epsilon_F)]^2}\right]
\end{align}
\subsection{Molecule renormalization} 
\begin{align}
\partial_k \mphi&= \frac{h_k^2 k }{2 \pi } \frac{\Theta\left(k^2 + \epsilon_F + \frac{\md}{\Ad}\right) }{\Ad(2 k^2 + \epsilon_F )+ \md}- \frac{\lambda k}{2 \pi } \Theta(\epsilon_F - k^2) \\
\partial_k \Aphi&=- \frac{h_k^2 k \Ad }{2 \pi } \frac{\Theta\left(k^2 + \epsilon_F + \frac{\md}{\Ad}\right) }{[\Ad(2 k^2 + \epsilon_F )+ \md]^2}
\end{align}

\subsection{Three-body renormalization}
\subsubsection{Bubble}
\begin{align}
A_1=\frac{\lambda_k^2 k}{\pi}\left[\frac{\Theta \left( k^2+\frac{2 \mphi}{\Aphi}+\epsilon_F \right)}{  \Aphi \left(3 k^2+\epsilon_F \right)+2 \mphi}+ \frac{\Theta \left(k^2-\frac{2 \mphi}{\Aphi}-\epsilon_F\right)\Theta\left( \epsilon_F -2 k^2\right)}{3 \Aphi k^2-\Aphi \epsilon_F -2 \mphi}-\frac{\Theta \left(-\frac{k^2}{2}-\frac{\mphi}{\Aphi}\right)\Theta \left(\epsilon_F -2 k^2\right)}{3 \Aphi k^2-2
		\Aphi \epsilon_F +2 \mphi}\right]
\end{align}
\subsubsection{Triangle}
\begin{align}
B_1&= \frac{2 h_k^2 k \lambda_k  \Theta \left(k^2+\frac{2 \mphi}{\Aphi}+\epsilon_F \right)}{\pi  \left(\Ad \left(2 k^2+\epsilon_F \right)+\md \right) \left(\Aphi \left(3 k^2+\epsilon_F \right)+2 \mphi\right)}\\
B_2&= -\frac{2 \Ad h_k^2 k \lambda_k  \Theta \left(-\frac{2 \mphi+\Aphi \left(k^2+\epsilon_F \right)}{2 \Aphi}\right)}{\pi 
	\left(\Ad \left(2 k^2+\epsilon_F \right)+\md\right) \left(-2 \Ad \mphi+\Ad \Aphi \left(k^2+\epsilon_F
	\right)+2 \Aphi \md\right)}\\
B_3&= -\frac{4 \Aphi h_k^2 k \lambda_k  \Theta \left(-\frac{k^2}{2}-\frac{\mphi}{\Aphi}\right)\Theta\left(\epsilon_F -2 k^2\right)}{\pi  \left(3 \Aphi
	k^2-2 \Aphi \epsilon_F +2 \mphi\right) \left(-2 \Ad \mphi+\Ad \Aphi k^2+2 \Aphi \md\right)}\\
B_4&= \frac{4 \Aphi h_k^2 k \lambda_k  \Theta \left(k^2-\frac{2 \mphi}{\Aphi}-\epsilon_F\right) \Theta\left(\epsilon_F -2 k^2\right)}{\pi  \left(\Aphi
	\left(\epsilon_F -3 k^2\right)+2 \mphi\right) \left(2 \Ad \mphi+\Ad \Aphi \left(k^2-\epsilon_F \right)-2 \Aphi
	\md\right)}
\end{align}
\subsubsection{Square}
\begin{align}
C_1&= \frac{h_k^4 k \Theta \left(k^2+\frac{2 \mphi}{\Aphi}+\epsilon_F \right)}{\pi  \left(\Ad \left(2 k^2+\epsilon_F \right)+\md \right)^2 \left(\Aphi \left(3 k^2+\epsilon_F \right)+2 \mphi\right)}\\
C_2&= \frac{ \Ad h_k^4 k  \Theta \left(-\frac{2 \mphi+\Aphi \left(k^2+\epsilon_F \right)}{2 \Aphi}\right)\left(2 \Ad \mphi - \Aphi\left[ 5 \Ad k^2 + 4 \md + 3 \Ad \epsilon_F \right] \right)}{\pi 
	\left(\Ad \left(2 k^2+\epsilon_F \right)+\md\right)^2 \left(-2 \Ad \mphi+\Ad \Aphi \left(k^2+\epsilon_F
	\right)+2 \Aphi \md\right)^2}\\
C_3&= -\frac{4 \Aphi^2 h_k^4 k   \Theta \left(-\frac{k^2}{2}-\frac{\mphi}{\Aphi}\right)\Theta\left(\epsilon_F -2 k^2\right)}{\pi  \left(3 \Aphi
	k^2-2 \Aphi \epsilon_F +2 \mphi\right) \left(-2 \Ad \mphi+\Ad \Aphi k^2+2 \Aphi \md\right)^2}\\
C_4&=- \frac{4 \Aphi^2 h_k^4 k \Theta \left(k^2-\frac{2 \mphi}{\Aphi}-\epsilon_F \right) \Theta\left(\epsilon_F -2 k^2\right)}{\pi  \left(\Aphi
	\left(\epsilon_F -3 k^2\right)+2 \mphi\right) \left(2 \Ad \mphi+\Ad \Aphi \left(k^2-\epsilon_F \right)-2 \Aphi
	\md\right)^2}
\end{align}
\subsubsection{Total}
\begin{align}
\partial_k \lambda= - \frac{\lambda_k^2}{h_k^2}\partial_k \md+ A_1+ \sum_{i=1}^4 B_i + \sum_{i=1}^4 C_i
\end{align}

\subsection{Fermion renormalization}
\begin{align}
\partial_k \mup&=\frac{h_k^2 k}{\Aphi \Ad \pi} \int_{-\pi}^{\pi} \frac{d\theta}{2\pi} \frac{\Theta\left(- k^2 - \frac{2 \mphi}{\Aphi}\right)\Theta\left(p^2 - 2 k p \cos(\theta)\right)}{-k^2 +  \frac{2\mphi}{\Aphi}-  \frac{2 \md}{\Ad}- 2 p^2 + 4 k p \cos(\theta) }\Bigg|_{p =\sqrt{\epsilon_F}}\nnl
&+\frac{h_k^2 k}{\Aphi \Ad \pi} \int_{-\pi}^{\pi} \frac{d\theta}{2\pi} \frac{\Theta\left(- k^2 - \frac{2 \mphi}{\Aphi}- p^2 - 2 k p \cos(\theta)\right)\Theta\left(p^2 + 2 k p \cos(\theta)\right)}{-k^2 +  \frac{2\mphi}{\Aphi}-  \frac{2 \md}{\Ad}+ p^2 + 2 k p \cos(\theta) }\Bigg|_{p =\sqrt{\epsilon_F}}\nnl
&-\frac{\lambda_k k }{2 \pi A_{\phi}} \Theta\left(-k^2-  \frac{2\mphi}{\Aphi}\right)\\
\partial_k \Aup&=-\frac{2 h_k^2 k}{\Aphi \Ad \pi} \int_{-\pi}^{\pi} \frac{d\theta}{2\pi} \frac{\Theta\left(- k^2 - \frac{2 \mphi}{\Aphi}\right)\Theta\left(p^2 - 2 k p \cos(\theta)\right)}{\left(-k^2 +  \frac{2\mphi}{\Aphi}-  \frac{2 \md}{\Ad}- 2 p^2 + 4 k p \cos(\theta)\right)^2 }\Bigg|_{p =\sqrt{\epsilon_F}}\nnl
&-\frac{2h_k^2 k}{\Aphi \Ad \pi} \int_{-\pi}^{\pi} \frac{d\theta}{2\pi} \frac{\Theta\left(- k^2 - \frac{2 \mphi}{\Aphi}- p^2 - 2 k p \cos(\theta)\right)\Theta\left(p^2 + 2 k p \cos(\theta)\right)}{\left(-k^2 +  \frac{2\mphi}{\Aphi}-  \frac{2 \md}{\Ad}+ p^2 + 2 k p \cos(\theta) \right)^2}\Bigg|_{p =\sqrt{\epsilon_F}}
\end{align}
\subsection{Effective potential}
\begin{align}
\partial_k U_k&= \frac{1}{4\pi } \left(\partial_k \frac{\mphi}{\Aphi} \right) \min\left(k^2, -2\frac{ \mphi}{\Aphi} \Theta\left(-\frac{\mphi}{\Aphi}\right)\right)\nnl
&+ \frac{1}{4\pi} \left(\partial_k \frac{\mup}{\Aup}\right) \left(\max\left(\epsilon_F- \frac{\mup}{\Aup},0\right)- \max\left(\epsilon_F- \frac{\mup}{\Aup}-k^2,0\right)\right)
\end{align}
\end{widetext}

\end{document}